\documentclass{emulateapj}
\usepackage{natbib}
\usepackage{color}

\shorttitle{One Hundred First Stars}  %===[not more than 44 characters]
\shortauthors{Hirano et al.}

\begin{document}

\title{{\it One Hundred First Stars} : Protostellar Evolution and the Final Masses}

\author{Shingo Hirano\altaffilmark{1},
Takashi Hosokawa\altaffilmark{2},
Naoki Yoshida\altaffilmark{3,4},
Hideyuki Umeda\altaffilmark{1}, \\
Kazuyuki Omukai\altaffilmark{5},
Gen Chiaki\altaffilmark{3},
and 
Harold W. Yorke\altaffilmark{6}}

\altaffiltext{1}{Department of Astronomy, School of Science, University of Tokyo, Bunkyo, Tokyo 113-0033, Japan}
\altaffiltext{2}{Department of Physics and Research Center for the Early Universe, University of Tokyo, Bunkyo, Tokyo 113-0033, Japan}
\altaffiltext{3}{Department of Physics, School of Science, University of Tokyo, Bunkyo, Tokyo 113-0033, Japan}
\altaffiltext{4}{Kavli Institute for the Physics and Mathematics of the Universe (WPI), Todai Institutes for Advanced Study, the University of Tokyo, Kashiwa, Chiba 277-8583, Japan}
\altaffiltext{5}{Astronomical Institute, Tohoku University, Sendai, Miyagi 980-8578, Japan}
\altaffiltext{6}{Jet Propulsion Laboratory, 
California Institute of Technology, Pasadena CA 91109, USA}

\setcounter{footnote}{-1}

\begin{abstract}  %===[not more than 250 words]

We perform a large set of radiation hydrodynamics simulations of primordial star 
formation in a fully cosmological context. 
\let\thefootnote\relax\footnote{\copyright 2013: All rights reserved}
Our statistical sample of {\it 100 First Stars} show that the first generation of stars 
have a wide mass distribution $M_{\rm popIII} = 10 \sim 1000 \ {\rm M_{\odot}}$. 
We first run cosmological simulations to generate a set of primordial star-forming gas clouds. 
We then follow protostar formation in each gas cloud and the subsequent 
protostellar evolution until the gas mass accretion onto the protostar 
is halted by stellar radiative feedback. 
The accretion rates differ significantly among the primordial gas clouds 
which largely determine the final stellar masses.
For low accretion rates the growth of a protostar is
self-regulated by radiative feedback effects and the final mass
is limited to several tens of solar masses.
At high accretion rates the protostar's outer envelope continues to expand and 
the effective surface temperature remains low; such protostars do not exert strong 
radiative feedback and can grow in excess to one hundred solar masses. 
The obtained wide mass range suggests that the first stars play a variety of roles 
in the early universe, by triggering both core-collapse supernovae and pair-instability 
supernovae as well as by leaving stellar mass black holes.
We find certain correlations between the final stellar mass and the physical 
properties of the star-forming cloud.
These correlations can be used to estimate the mass of the first star 
from the properties of the parent cloud or of the host halo,
without following the detailed protostellar evolution.

\end{abstract}

  %=== [A maximum of six subject keywords]

\section{Introduction}

The first stars played a key role in the early universe. 
Their first light initiated cosmic reionization, and the heavy elements synthesized by them enabled the formation of subsequent ordinary stellar populations 
\citep[e.g.,][]{alvarez06, johnson07, maio11, wise12, karlsson13}. 
The first stars could have also seeded the formation of supermassive 
black holes observed at high redshifts \citep{li07, volonteri12, hosokawa12a, hosokawa13}.
Understanding the formation of the first stars is thus necessary 
to understand the early history of the universe \citep[e.g.,][]{bromm09}.
Theoretical studies are rapidly advancing our knowledge on
how the first stars are formed from the cosmic primordial gas, 
supplemented by a variety of observations ranging 
from Galactic metal-poor stars to high redshift galaxies that
provide invaluable information on early star formation
\citep{bromm11}.
However, there still remain an important question
on the nature of the first generation of stars:
{\it what is the characteristic mass distribution
of the first stars?}   
The fate of a star and the strength of various stellar 
feedback effects on the local environment are largely determined by 
the star's mass,
and it is thus crucial to know the characteristic mass
of the first stars. 

The initial conditions of the first stars' formation are 
determined cosmologically; small density perturbations left 
over from the Big Bang drive, via gravitational instability, 
the formation of primordial star-forming gas clouds.
It has become possible to perform 
{\it ab initio} numerical simulations to follow this process
\citep[e.g.,][]{abel02, yoshida08, greif12}. 
Such cosmological simulations show that small dark matter halos 
($\sim 10^5$ - $10^6 \ {\rm M_{\odot}}$) 
forming at redshift $z \sim 20 - 30$ 
are cradles of the first stars \citep[e.g.,][]{yoshida03a, gao07, o'shea07}. 
Typically, a massive star-forming gas cloud of $\sim 1000 \ {\rm M_{\odot}}$, 
which is gravitationally unstable, is formed at the density peak of the host dark matter halo. 
The cloud gravitationally collapses 
with help of radiative cooling by hydrogen molecules 
\citep[e.g.,][]{matsuda69, palla83}. 
The collapse continues until 
a quasi-hydrostatic protostellar core ($\sim 0.01 \ {\rm M_{\odot}}$) is formed 
in the densest part of the cloud \citep[e.g.,][]{omukai98, yoshida08}.

There is still a long way for this protostar to 
grow in mass and ultimately reach the Pop III main-sequence.
The initially tiny protostellar core grows via accretion of the 
surrounding gas. 
The typical accretion rate in a primordial gas cloud is
$10^{-3} - 10^{-2} \ {\rm M_{\odot} \ yr^{-1}}$, 
with which most of the gas in the cloud can be accreted onto the star
during its stellar lifetime ($\sim$ Myr).
It has been postulated that 
the final stellar mass is set by the mass of the natal cloud,  
which is as massive as $M_{\rm cloud} \sim 100 - 1000 \ {\rm M_{\odot}}$.
This is too naive an estimate, however. 
The final stellar mass is determined by
a complex interplay between the growing
central protostar and the accreting gas.
Radiation from the protostar 
critically affects the process and possibly halts gas accretion
hence its mass growth \citep{hosokawa11, hosokawa12b}.

The strength of the UV radiative feedback depends on 
the stellar luminosity and effective temperature, 
which are determined by the evolution of the protostar itself.
The structure and evolution of an accreting primordial protostar have been 
studied in detail by numerically solving the interior structure 
\citep[e.g.,][]{stahler86a, omukai01, omukai03a, ohkubo09, hosokawa12a}. 
\cite{mckee08} develop an analytic model of the protostar's evolution 
and apply it to a particular case to find 
the final stellar mass to be around $140 \ {\rm M_{\odot}}$.

Numerical simulations are necessary to follow the growth of the protostar 
and the detailed interplay between the star and the surrounding gas
self-consistently,
because the UV feedback influences directly and almost instantaneously
the mass accretion rate.
The first such simulations of \citet{hosokawa11} show that 
the UV radiative feedback is strong enough to halt gas accretion 
when the stellar mass is around a few tens of solar masses 
\citep[see also][]{stacy12, susa13}.

Observationally, the typical mass of the first stars can be inferred,
for example, from the elemental abundance patterns of metal-poor stars 
in the Galaxy \citep{caffau11}.
Interestingly, the abundance patterns found in the atmospheres of
a few extremely metal-poor stars can be reconciled, 
if the metals are produced in supernovae of 
progenitor stars with a mass $\le 100 \ {\rm M_{\odot}}$ \citep{umeda05}. 
This relative small mass is consistent with the 
conclusion of recent theoretical studies on primordial star formation \citep{hosokawa11}.
Very massive stars with $150 - 300 \ {\rm M_{\odot}}$ end their lives 
as pair-instability supernovae \citep[PISNe;][]{barkat67, bond84, heger02}.
PISNe would imprint a peculiar abundance pattern in metal-poor stars,
a pattern for which there is no clear observational indication
\citep{tumlinson04, frebel09}. 

Previous theoretical studies examine 
only a limited number of cases to derive the final stellar masses. 
The characteristic mass and the overall mass distribution
are yet unknown.
A large set of simulations beginning from the earliest phases of
structure formation in the universe are needed
to derive the mass distribution of the first stars.

In this paper we study the statistical properties of the first stars,
using more than one hundred clouds produced in a cosmological simulation.
To this end, we first run a large-volume cosmological simulation which follows
the formation of primordial clouds in dark matter halos and
their initial run-away collapse \citep{yoshida06, yoshida08, hirano13}. 
We then switch to the radiation hydrodynamic simulations 
coupled with stellar evolution calculations, 
which follow the subsequent evolution in the accretion phase 
until the mass accretion onto the protostar is shut off 
by strong stellar UV feedback \citep{hosokawa11, hosokawa12b}. 
From these simulations we obtain a distribution of stellar masses,
which, as we find, ranges from just under ten to more than one thousand solar masses.

The rest of the paper is organized as follows. 
In Section 2, we briefly describe 
the numerical methods employed in our cosmological simulation, the 
radiation hydrodynamic simulations, and the stellar evolution calculations. 
Section 3 presents the results of our {\it 100 First Stars} simulations; 
statistical variations regarding the first stars' formation and 
the resulting distribution of stellar masses.
In Section 4, we investigate the origins of 
the wide distribution of the final stellar masses.
Section 5 and 6 provide the relevant discussions and concluding remarks of this study.

\section{Methods}

We first perform cosmological $N$-body SPH (Smoothed-Particle-Hydrodynamics) simulations 
of early structure formation. We locate and label dense and cold gas clumps hosted
by small mass dark halos as sites for primordial star formation.
We follow the pre-stellar collapse of each cloud up to the formation of a protostar.
We then switch to two dimensional axisymmetric radiation hydrodynamics simulations
of protostellar evolution that start with the output of
our cosmological simulations.
The following subsections describe our numerical methods. 
We highlight several novel techniques that are introduced for
the present study.

%=== TABLE1 ===%
\begin{table*}
\begin{center}
\caption{Parameters of Cosmological Simulations\label{t1}}
\begin{tabular}{rrcrr@{.}llc}
\tableline\tableline\noalign{\smallskip}
$N_{\rm sample}$ & $L_{\rm box}$ & $N_{\rm zoom}$ & $l_{\rm soft,zoom}$ & \multicolumn{2}{c}{$m_{\rm DM,zoom}$} & $m_{\rm Gas,zoom}$ & \\ 
 & $({\rm h^{-1} \ kpc})$ & & $({\rm h^{-1} \ pc})$ & \multicolumn{2}{c}{$({\rm h^{-1} \ M_{\odot}})$} & $({\rm h^{-1} \ M_{\odot}})$ & \\
\noalign{\smallskip}\tableline\noalign{\smallskip}
 3 &   50 & 1024 &  1.0 &  0 & 0142 & 0.00288 & (1) \\
 3 &  150 & 2048 &  1.5 &  0 & 0478 & 0.00971 & (1) \\
 1 &  300 & 2048 &  2.9 &  0 & 383  & 0.0772  & (1) \\
 7 & 1000 & 3072 &  6.5 &  2 & 16   & 0.436   & (2) \\
75 & 1000 & 1536 & 13.5 & 17 & 3    & 3.49    & (3) \\
21 & 2000 & 3072 & 13.5 & 17 & 3    & 3.49    & (2) \\
\noalign{\smallskip}\tableline\noalign{\smallskip}
\end{tabular}
\tablecomments{
Column 1: Number of star-forming clouds, 
Column 2: Simulation box size (comoving), 
Column 3: The effective grid resolution in the zoomed region, 
Column 4: Gravitational softening length (comoving), 
Column 5: Dark matter particle mass,
Column 6: Gas particle mass, and
Column 7: Reference of the adopted cosmological parameters.
}
\tablerefs{
(1) Larson et al. 2011; 
(2) Komatsu et al. 2009; 
(3) Komatsu et al. 2011.}
\end{center}
\end{table*}

%%%
%\begin{deluxetable}{rrcrr@{.}llc}
%\tablewidth{0pt}
%\tablenum{1}
%\tablecaption{Parameters of Cosmological Simulations}
%\tablehead{
%\colhead{$N_{\rm sample}$} & 
%\colhead{$L_{\rm box}$} & 
%\colhead{$N_{\rm zoom}$} & 
%\colhead{$l_{\rm soft,zoom}$} &
%\multicolumn{2}{c}{\colhead{$m_{\rm DM,zoom}$}} & 
%\colhead{$m_{\rm Gas,zoom}$} & 
%\colhead{} \\ 
%\colhead{} & 
%\colhead{$({\rm h^{-1} \ kpc})$} & 
%\colhead{} & 
%\colhead{$({\rm h^{-1} \ pc})$} &
%\multicolumn{2}{c}{\colhead{$({\rm h^{-1} \ M_{\odot}})$}} & 
%\colhead{$({\rm h^{-1} \ M_{\odot}})$} & 
%\colhead{}}
%\startdata\
% 3 &   50 & 1024 &  1.0 &  0 & 0142 & 0.00288 & (1) \\
% 3 &  150 & 2048 &  1.5 &  0 & 0478 & 0.00971 & (1) \\
% 1 &  300 & 2048 &  2.9 &  0 & 383  & 0.0772  & (1) \\
% 7 & 1000 & 3072 &  6.5 &  2 & 16   & 0.436   & (2) \\
%75 & 1000 & 1536 & 13.5 & 17 & 3    & 3.49    & (3) \\
%21 & 2000 & 3072 & 13.5 & 17 & 3    & 3.49    & (2) \\
%\enddata
%\tablecomments{
%Column 1: Number of star-forming clouds, 
%Column 2: Simulation box size (comoving), 
%Column 3: The effective grid resolution in the zoomed region, 
%Column 4: Gravitational softening length (comoving), 
%Column 5: Dark matter particle mass,
%Column 6: Gas particle mass, and
%Column 7: Reference of the adopted cosmological parameters.
%}
%\tablerefs{
%(1) Larson et al. 2011; 
%(2) Komatsu et al. 2009; 
%(3) Komatsu et al. 2011.}
%\label{t1}
%\end{deluxetable}
%%%

\subsection{Cosmological Simulations: Primordial Gas Clouds}

Our $\Lambda$-Cold Dark Matter ($\Lambda$CDM) 
simulations adopt
the cosmological parameters consistent with the WMAP 9-year data \cite{komatsu11}. 
The basic simulation parameters are summarized in Table \ref{t1}.
All the cosmological simulations are initialized at $z_{\rm ini} = 99$.

We use the parallel $N$-body / SPH solver 
$GADGET$-2 \citep{springel05} in its version 
suitably adopted for primordial star formation
as in \cite{hirano13}.
We solve chemical rate equations for 14 primordial species
(${\rm e^-}$, ${\rm H}$, ${\rm H^+}$, ${\rm H^-}$, ${\rm He}$, 
${\rm He^+}$, ${\rm He^{++}}$, ${\rm H_2}$, ${\rm H_2^+}$, ${\rm D}$, 
${\rm D^+}$, ${\rm HD}$, ${\rm HD^+}$, ${\rm HD^-}$)
as in \cite{yoshida06, yoshida07}.
To accurately evaluate trapping effects of cooling radiation,
we employ the Sobolev method for ${\rm H_2}$ line cooling and 
a ray-tracing method for continuum cooling by ${\rm H_2}$ 
collision-induced emission \citep{hirano13}.

The main parent simulations have a sufficiently large volume of 
$L_{\rm box}$ = 1 and 2 ${\rm h^{-1} \ Mpc}$ on a side. 
We also use a few additional simulations with smaller volumes. 
There are a number of primordial clouds at redshifts $10 - 30$. 
The typical mass of the star-forming clouds, 
$\sim 1000 \ {\rm M_{\odot}}$, 
is resolved with more than 128 SPH particles. 
We employ a hierarchical zoom-in re-simulation method
to resolve a primordial gas cloud with a progressively 
larger number of particles.
Each refined region has a spherical shape of 
radius $130 \ {\rm h^{-1} \ kpc}$ in the case of 
$L_{\rm box} = 1 \ {\rm h^{-1} \ Mpc}$. 
All the re-simulations are started from z = 99.

We run a friends-of-friends halo finder to locate 
dense, clustered regions in the parent cosmological simulations. 
Then we generate the initial conditions for the selected regions, 
where the mass- and spatial-resolution are increased. 
The corresponding small-scale density fluctuations are added 
suitably as given by our adopted $\Lambda$CDM cosmology.
The selection of the gas clouds is done as follows.
First, we select a virialized dark matter halo whose mean density 
within the virial radius is 200 times greater than the mean density of the universe, 
i.e., $n_{\rm dm} > n_{\rm vir} \sim 200 \ n_{\rm univ}$. 
Then, if the halo is the densest structure within a 2 physical ${\rm kpc}$ radius 
around itself, we assume the halo has not been influenced 
by supernova explosions of nearby stars \citep[e.g.,][]{ritter12}
nor by radiative feedback \citep[e.g.,][]{o'shea08, agarwal12, johnson13}.
Primordial gas clouds forming at the centers of such halos 
are thought to bear the first generation of stars. 
In this way, we locate and select a sample of 110 gas clouds 
from the parent cosmological simulations.

Our zoom-in simulations follow the formation and gravitational run-away 
collapse of the selected primordial gas clouds. 
We use the particle-splitting technique of \cite{kitsionas02} 
to achieve a wide dynamic range. 
By this method, the local Jeans length is always resolved by 
10 times the local smoothing length of the SPH particles.
We stop our zoom-in simulations
when the central density of the cloud reaches 
$n_{\rm H,cen} = 10^{13} \ {\rm cm^{-3}}$.
At this moment, the mass of the lightest gas particle is 
$m_{\rm gas} \sim 10^{-5} \ {\rm M_{\odot}}$, and 
thus the nominal mass resolution is $\sim 0.01 \ {\rm M_{\odot}}$. 
The model outputs at this time are used to generate the {\it initial}
conditions for our protostellar evolution calculations.

\subsection{Radiation Hydrodynamic Simulations: Evolution of 
Accreting Protostars}

We follow the evolution of the protostars in the accretion phase 
using 2D axisymmetric radiation hydrodynamic (RHD) calculations 
coupled with the stellar structure evolution \citep{hosokawa11}.
The nested-grid method \citep[e.g.,][]{yorke95, yorke99, yorke02} 
is employed in order to achieve an extremely wide dynamic range.  
Our calculations utilize 9 level hierarchical grids, whereby 
the coarsest grid size is $\sim 6400 \ {\rm AU}$ and 
the finest is $\sim 25 \ {\rm AU}$ in a $\sim 1$ pc volume.
Data from the three-dimensional cosmological simulation 
are mapped onto two-dimensional axisymmetric meshes as follows.
First, we fix the principal rotational axis by calculating 
the average angular momentum vector of 
the gas within a 0.01 pc region around the cloud center.
Then we take averages of physical quantities along the azimuthal 
direction around the rotational axis under the 8 level hierarchical grids.
The free flow boundary conditions are adopted in the RHD simulations; 
the material can escape and enter through the edges of the simulation domains. 
We assume that the density at the boundary is the same as 
that of the outermost grid. 
To eliminate the effects of this boundary condition, 
we set a sufficiently large box size 
so that the inflow cannot affect the accretion onto the central protostar.
The computational domain is $1.2 \ {\rm pc}$ on a side, which
typically contains a few thousands solar masses of gas.

We consistently follow the structure and evolution of the
central accreting star and the hydrodynamics of the accreting gas
which is irradiated by the stellar radiation as in \cite{hosokawa11}. 
The mass accretion rate onto the protostar is directly obtained from 
the inflow rate of gas into the central sink cell in the RHD simulation. 
Our code treats non-equilibrium chemistry and radiative
processes in a primordial gas \citep[summary of the reactions are found in][]{hosokawa11}. 
We have verified that HD cooling, which
sometimes operates in the pre-collapse stage (see Sec. 3.2.2 below), 
is unimportant in the vicinity of the protostar and in the accretion disk. 
We switch off the HD chemistry in our protostellar evolution 
calculations \citep{hosokawa12b}.

The evolution of the protostar is calculated 
by numerically solving the interior structure with this accretion rate 
\citep[e.g.,][]{omukai03a, hosokawa09a}. 
The stellar luminosity and the effective temperature are provided 
by the stellar model.
The spectral energy distribution is calculated from these basic
 properties of the star.

We explore various cases with different gas accretion rates
and different protostellar evolution.
To this end, we have added several modifications to the code 
used in \cite{hosokawa11, hosokawa12b}.
The improvements are summarized as follows (see also Appendixes A and B).

\begin{itemize}
\item 
Angular momentum transport in a rapidly accreting circumstellar disk 
is driven mainly by its non-axisymmetric structures such as spiral arms. 
\cite{hosokawa11} adopt the so-called $\alpha$-viscosity model \citep{shakura73} 
using a constant gravo-viscous parameter $\alpha$ 
everywhere on the equator of the disk \citep[e.g.,][]{yorke02}. 
Recent 3D numerical simulations suggest that, in a self-gravitating disk, 
the effective $\alpha$ can be estimated as a function of 
the Toomre $Q$-parameter \citep{zhu10a, takahashi13}. 
Motivated by this, we adopt a functional form proposed by \cite{gammie96}, 
and allow $\alpha$ to spatially 
vary with the Toomre $Q$-parameter evaluated at each cell on the equator, 
$\alpha(Q(r))$. Details of this procedure are described in Appendix A.
\item In a number of our cases we find rather high accretion rates, 
exceeding 0.01 ${\rm M_{\odot} \ yr^{-1}}$ (Sec. 3.3 below).
For sufficiently high accretion rates, the protostellar evolution differs significantly
from the fiducial case, whereby the star contracts to reach the zero-age main-sequence 
\citep[][also see Sec. 2.2.1]{omukai03a, hosokawa12a}. 
In practice, we occasionally encounter convergence difficulties with time-dependent 
accretion histories resulting from our RHD simulations. 
We avoid this technical problem by employing an analytic model of stellar evolution. 
In essence, we assume that the rapidly accreting star, whose envelope experiences
expansion and oscillations, nevertheless evolves on a certain averaged
track on the mass-radius plane. We describe the details of our analytic model in Appendix B.
\end{itemize}

We assume that the mass accretion and hence the growth of the central star 
are halted due to stellar radiative feedback, 
when the mass accretion rate falls below $10^{-4} \ {\rm M_{\odot}} \ {\rm yr^{-1}}$. 
We stop the calculation at this time. 
In fact \cite{hosokawa11, hosokawa12a} show that 
the accretion rate continues to decrease after that and further stellar growth in mass is negligible.
The final stellar mass generated by each of the 110 collapsing clouds represent a
statistical sample of the final masses of the first stars, $M_{\rm popIII}$.

\subsubsection{Three Paths of the Protostellar Evolution}

It is known that an accreting protostar goes through a qualitatively 
different evolution 
depending on the accretion rate \citep[e.g.,][]{omukai03a, hosokawa12a}.
We classify the notably different protostellar evolutionary paths 
into three characteristic cases.
For completeness we first review briefly each evolutionary path, in order 
to better describe our main results presented in the following sections.
In the figures hereafter, the three evolutionary paths are distinguished 
by using different colors (red, blue, and black) for clarity.

\paragraph{KH Contracting Protostar (P1; $\dot{M} < 4 \times 10^{-3}$ ${\rm M_{\odot} \ yr^{-1}}$)} \

The evolution of an accreting protostar is approximately determined
by the balance between the following two physical timescales. 
One is the stellar Kelvin-Helmholtz (KH) timescale, 
over which the protostar radiates away its gravitational energy,
\begin{equation}
t_{\rm KH} \equiv \frac{G M_{\rm star}^2}{R_{\rm star} L_{\rm star}} \ , 
\label{eq:KH-timescale}
\end{equation}
where $M_{\rm star}$, $R_{\rm star}$, and $L_{\rm star}$ are the stellar mass, 
radius, and luminosity, respectively.
This is an approximate timescale for a non-accreting protostar 
to contract to the main-sequence.
The other is the accretion timescale, over which the protostar
doubles its mass by accretion,
\begin{equation}
t_{\rm acc} \equiv \frac{M_{\rm star}}{\dot{M}} \ ,
\label{eq:ACC-timescale}
\end{equation}
where $\dot{M}$ is the mass accretion rate.

\begin{figure}
\begin{center}
\resizebox{7cm}{!}{\includegraphics[clip]{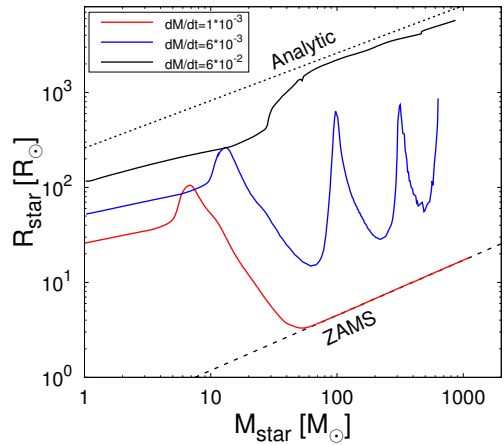}}
\caption{
Evolution of the accreting protostar's radius 
for three constant accretion rates
$\dot{M} = 1 \times 10^{-3}$, $6 \times 10^{-3}$, and 
$6 \times 10^{-2} \ {\rm M_{\odot} \ yr^{-1}}$ 
\citep[taken from][]{hosokawa12a}.
The dotted and dashed lines represent the mass-radius relations for 
super-giant protostars (Eq. \ref{eq:SGPS-radius}) and ZAMS stars.
}
\label{plot:3path_woUV}
\end{center}
\end{figure}

In Figure \ref{plot:3path_woUV}, the case with $\dot{M} = 10^{-3} \ {\rm M_{\odot} \ yr^{-1}}$ 
corresponds to the evolution of the KH contracting protostar.
In an early evolutionary stage when $M_{\rm star} \le 5 \ {\rm M_{\odot}}$, 
the KH timescale is longer than the accretion timescale, $t_{\rm KH} \gg t_{\rm acc}$, 
so that the energy deposition by accretion is faster than the energy loss by radiation 
\citep[this is called ``adiabatic accretion'';][]{stahler86a}.
When $M_{\rm star}$ exceeds $5 \ {\rm M_{\odot}}$ the stellar luminosity 
begins to increase sharply, and 
the KH timescale quickly decreases with increasing stellar mass. 
The KH timescale eventually falls below the accretion timescale, 
and the protostar is able to efficiently radiate away its internal energy; it
begins to contract ($M_{\rm star} \ge 7 \ {\rm M_{\odot}}$; ``KH contraction''). 
KH contraction continues until the star begins core hydrogen burning
at $M_{\rm star} \simeq 40 \ {\rm M_{\odot}}$, a point at which it has 
essentially reached the zero-age main-sequence (ZAMS).

\paragraph{Oscillating Protostar (P2; $\dot{M} > 4 \times 10^{-3}$ ${\rm M_{\odot} \ yr^{-1}}$)} \

\cite{omukai03a} show that, with rapid mass accretion, 
the star begins expanding before reaching the ZAMS.
Figure \ref{plot:3path_woUV} illustrates this for the case 
$\dot{M} = 6 \times 10^{-3} \ {\rm M_{\odot} \ yr^{-1}}$.
Abrupt expansion occurs when the stellar total luminosity approaches 
the Eddington value during KH contraction.
The critical accretion rate above which this abrupt expansion occurs has been
studied by \cite{omukai03a}.
The total luminosity at the accreting envelope 
\begin{equation}
L_{\rm env} = L_{\rm star} + L_{\rm acc} \simeq L_{\rm star} + \frac{GM_{\rm star}\dot{M}}{R_{\rm star}}
\label{eq:Lstar}
\end{equation}
becomes equal to the Eddington luminosity $L_{\rm Edd}$ 
when the star reaches the ZAMS. 
The condition for the critical value is thus 
$L_{\rm env}|_{\rm ZAMS} = L_{\rm ZAMS} + L_{\rm acc, ZAMS} = L_{\rm Edd}$, from which we obtain
\begin{equation}
L_{\rm Edd} = L_{\rm ZAMS} + \frac{GM_{\rm ZAMS}\dot{M}_{\rm P2}}{R_{\rm ZAMS}} \ ,
\label{eq:P2-criterion_at_Edd}
\end{equation}
\begin{eqnarray}
\dot{M}_{\rm P2} & = & \frac{R_{\rm ZAMS}}{GM_{\rm ZAMS}}(L_{\rm Edd} - L_{\rm ZAMS}) \nonumber \\
& \approx & 4 \times 10^{-3} \ {\rm M_{\odot}} \ {\rm yr^{-1}} \ .
\label{eq:P2-criterion}
\end{eqnarray}
Note that $\dot{M_{\rm P2}}$ given by this equation is a function of $M_{\rm star}$, 
but its dependence is weak \citep{omukai03a}.
If the mass accretion rate is higher than this critical value during KH contraction, 
$\dot{M} > \dot{M}_{\rm P2}$, 
the star's total luminosity approaches the Eddington value before arriving at the ZAMS. 
The star can not contract any more and begins to expand. 
This {\it bloating} results from the high specific entropy gain in surface layers, 
where the opacity is higher than the interior, as they absorb a part of the outward heat flux. 
However, only these surface layers begin to inflate; 
the bulk of the stellar mass continues KH contraction (see below). 
As the star inflates, the Eddington ratio ($L_{\rm tot} / L_{\rm Edd}$) 
decreases with decreasing accretion luminosity.  After reaching a 
significantly lower Eddington ratio
the star resumes KH contraction, thereby completing one cycle of oscillation.
The oscillatory timescale in this evolutionary stage is approximately given by
the KH timescale, the thermal adjustment timescale of the star.

\paragraph{Super-Giant Protostar (P3; $\dot{M} > 4 \times 10^{-2}$ ${\rm M_{\odot} \ yr^{-1}}$)} \

\cite{hosokawa12a} show that, if the accretion rate is higher than
\begin{equation}
\dot{M}_{\rm P3} \approx 4 \times 10^{-2} \ {\rm M_{\odot}} \ {\rm yr^{-1}} \ ,
\label{eq:P3-criterion}
\end{equation}
the protostar enters a third evolutionary path, 
whereby the stellar radius monotonically increases with stellar mass according to the relation
\begin{equation}
R_{\rm star} \simeq 2.6 \times 10^2 \ {\rm R_{\odot}} \ \left( \frac{M_{\rm star}}{{\rm M_{\odot}}} \right)^{1/2} 
\label{eq:SGPS-radius}
\end{equation}
for $M_{\rm star} > 100 \ M_{\odot}$. 
The above scaling is independent of the mass accretion rate
as long as it is greater than the critical value, $\dot{M}_{\rm P3}$. 
In Figure \ref{plot:3path_woUV}, 
the case with $6 \times 10^{-2} \ {\rm M_{\odot} \ yr^{-1}}$ demonstrates 
this behavior. 
Note that, even in this case, 
the timescale inversion to $t_{\rm KH} < t_{\rm acc}$ occurs 
at $M_{\rm star} \simeq 30 \ M_{\odot}$.
This means that most of the stellar interior contracts, radiating the energy away and
only a surface layer significantly inflates.
The star thus has a highly inhomogeneous structure, whereby
the contracting core is surrounded by a bloating envelope,
similar to red-giant stars.
This evolutionary path is appropriately called the ``super-giant protostar'' \citep{hosokawa12a}.

\cite{hosokawa12a} calculate the protostar evolution with large constant accretion rates,
with which the protostar evolves on the ``super-giant'' track.
In the present paper, we use varying mass accretion which is self-consistently calculated.
We find that protostars begin to contract when the accretion rates become sufficiently low.

\section{Results}

\subsection{Overview}

\begin{figure}
\begin{center}
\resizebox{6cm}{!}{\includegraphics[clip]{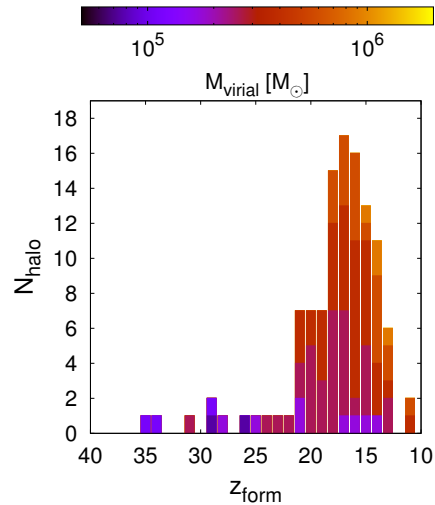}}
\caption{
The number of dark matter halos that host star-forming gas clouds.
The histogram shows the distribution of redshifts 
when the central gas density reaches $\sim 10^6 \ {\rm cm^{-3}}$. 
The histograms are colored according to the virial masses
using the color scale displayed at the top.
}
\label{plot_LIST_MpopIII_multi}
\end{center}
\end{figure}

\begin{figure*}
\begin{center}
\resizebox{10cm}{!}{\includegraphics[clip]{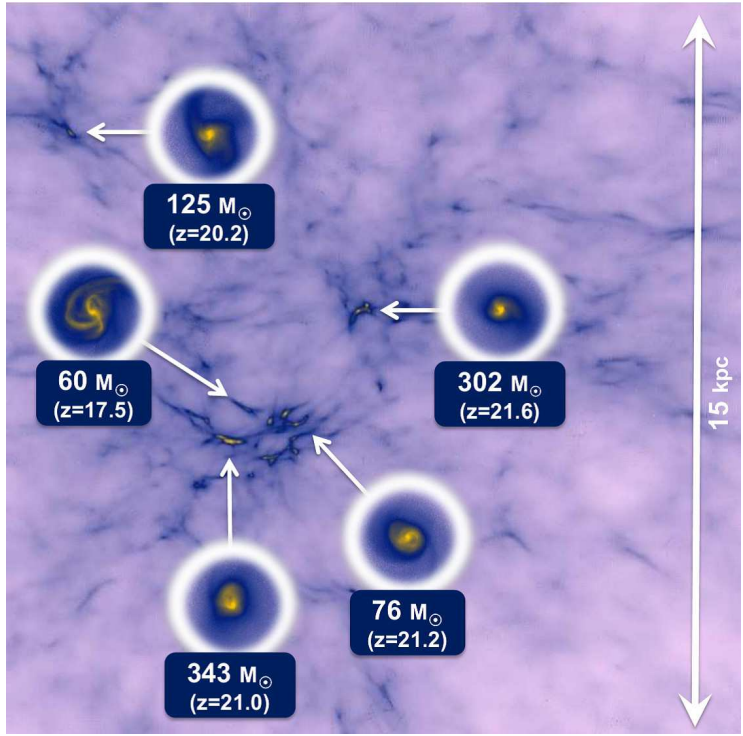}}
\caption{
Projected gas density distribution 
at $z = 25$ in one of our cosmological simulations.
We show five primordial star-forming clouds in a cube
of 15 kpc on a side. 
The circles show the zoom-in to the central 1 pc region of the clouds
at the respective formation epoch.
The masses of the first stars formed in these clouds are
$60$, $76$, $125$, $303$, and $343$ ${\rm M_{\odot}}$, respectively.
}
\label{f1}
\end{center}
\end{figure*}

The results presented here are based on a sequence of simulations
with different computer codes.
We first perform SPH cosmological simulations 
to follow the formation of the primordial gas clouds 
which gravitationally collapse in the center of dark matter halos. 
The histogram in Figure \ref{plot_LIST_MpopIII_multi} shows that 
our sample of 110 dark matter halos have a wide range of masses 
$M_{\rm virial} = 10^5 - 10^6 \ {\rm M_{\odot}}$
distributed over redshifts $z = 35 - 11$, most of which are at $z = 20 - 15$.
Figure \ref{f1} shows an example of the resulting gas density concentrations
arising in five such dark matter halos together
with insets of their zoomed-in structure, represented by white circles corresponding to 1 pc. 
As expected, the five clouds have different structures of density, velocity, and temperature.
The resulting stellar masses are also different as indicated in the figure.

\begin{figure*}
\begin{center}
\resizebox{10cm}{!}{\includegraphics[clip]{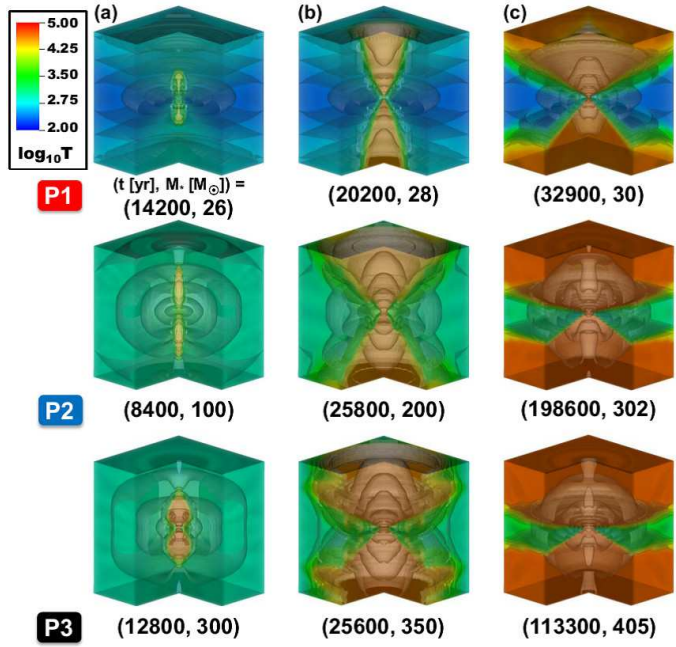}}
\caption{
Expanding H{\sc ii} regions around the primordial protostar 
for three in our sample of 110 clouds
(the same ones as in Figure \ref{plot_EXTEL_Mstr-dMdt-wowi}).
We show the structure and the evolution of the accreting gas 
from left to right. The plotted regions are cubes with
$60000 \ {\rm AU}$ on a side.
The colors indicate gas temperature and the
contours show the density structure.
The main accretion takes place 
through the accretion disk on the equatorial plane. 
As the central protostar becomes more massive and 
the surface temperature increases, 
the ionizing photon production of the central star 
increases. H{\sc ii} regions are launched
into the polar direction and the opening angles grow with time,
eventually stopping the accretion.
}
\label{f2}
\end{center}
\end{figure*}

\begin{figure}
\begin{center}
\resizebox{7cm}{!}{\includegraphics[clip]{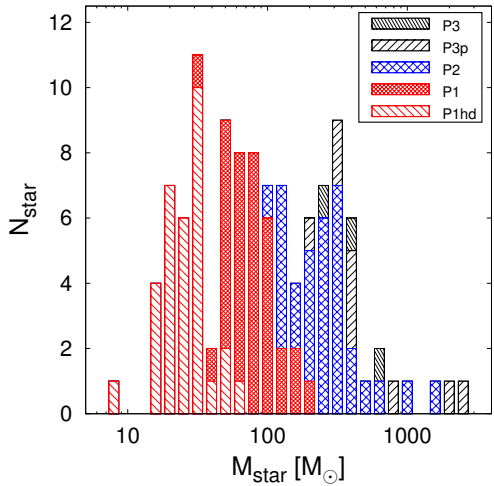}}
\caption{
The final distribution of the calculated stellar masses 
for our 110 first stars. 
The red, blue, and black histograms represent 
the different paths of protostellar evolution; 
P1: KH contracting protostar (red), 
P2: oscillating protostar (blue), and 
P3: super-giant protostar (black).
See text in Sec. 2.2.1 for details. 
P1hd refers to the cases in which the gas clouds are formed
by HD cooling and evolve on low-temperature tracks.
P3p (predicted) indicates the same cases as P3, except that
the final masses are calculated from a correlation between 
the properties of the cloud and the resulting stellar mass 
(Eq. \ref{eq:MpopIII-dMdtJeans}; see Appendix B).
}
\label{plot_IMF-PopIII_3PATHp}
\end{center}
\end{figure}

\begin{table}[t]
\begin{center}
\caption{Evolutionary Paths\label{t2}}
\begin{tabular}{cllc}
\tableline\tableline\noalign{\smallskip}
 & Path & $\dot{M}$ & $N_{\rm sample}$ \\
 & & $({\rm M_{\odot}} \ {\rm yr^{-1}})$ & \\
\noalign{\smallskip}\tableline\noalign{\smallskip}
P1 & KH Contracting Protostar & $<$ $0.004$ & 67 \\
P2 & Oscillating Protostar & $>$ $0.004^1$ & 31 \\
P3 & Super-Giant Protostar      & $>$ $0.04^2$  & 12 \\
\noalign{\smallskip}\tableline\noalign{\smallskip}
\end{tabular}
\end{center}
\tablecomments{
Column 2: Accretion rate for each path, and 
Column 3: The number of stars in our sample.
}
\tablerefs{
(1) Omukai \& Palla 2003;
(2) Hosokawa et al. 2012.
}
\end{table}
%%%
%\begin{deluxetable}{cllc}
%\tablewidth{0pt}
%\tablenum{2}
%\tablecaption{Evolutionary Paths}
%\tablehead{
%\colhead{} & 
%\colhead{Path} & 
%\colhead{$\dot{M}$} &
%\colhead{$N_{\rm sample}$} \\
%\colhead{} & 
%\colhead{} & 
%\colhead{$({\rm M_{\odot}} \ {\rm yr^{-1}})$} &
%\colhead{}} 
%\startdata
%P1 & KH Contracting Protostar & $<$ $0.004$ & 67 \\
%P2 & Oscillating Protostar & $>$ $0.004^1$ & 31 \\
%P3 & Super-Giant Protostar      & $>$ $0.04^2$  & 12 \\
%\enddata
%\tablecomments{
%Column 2: Accretion rate for each path, and 
%Column 3: The number of stars in our sample.
%}
%\tablerefs{
%(1) Omukai \& Palla 2003;
%(2) Hosokawa et al. 2012.}
%\label{t2}
%\end{deluxetable}
%%%

After the formation of a protostellar core at the center of the collapsing cloud, 
we switch to the 2D RHD calculations for each individual dark matter halo
and follow the evolution during the later accretion stages.
Figure \ref{f2} shows snapshots from three of our examined cases, 
which exemplify the three different evolutionary paths (P1, P2, and P3). 
We see that in each case a bipolar H{\sc ii} region forms
(Figure \ref{f2}a), which subsequently grows at varying rates 
as the stellar mass increases.
The mass accretion onto the protostar is finally shut off 
by the strong UV  radiative feedback 
caused by the dynamical expansion of the H{\sc ii} region 
\citep[Figure \ref{f2}c; see also][]{hosokawa11}. 
Figure \ref{plot_IMF-PopIII_3PATHp} shows 
the distribution of the final stellar masses obtained in our simulations 
(a summary is given in Table \ref{t2}). 
We see a large scatter of resulting stellar masses, 
ranging from 9.9 ${\rm M_{\odot}}$ to 1621 ${\rm M_{\odot}}$.
However, the bulk of them is distributed around 
several tens or a few hundreds of solar masses.
We study the origin of this distribution in Sec. 4 in detail.
Here, we merely note that the distribution of stellar masses 
does not mirror the distribution of dark matter halo masses.

\begin{figure}
\begin{center}
\resizebox{6cm}{!}{\includegraphics[clip]{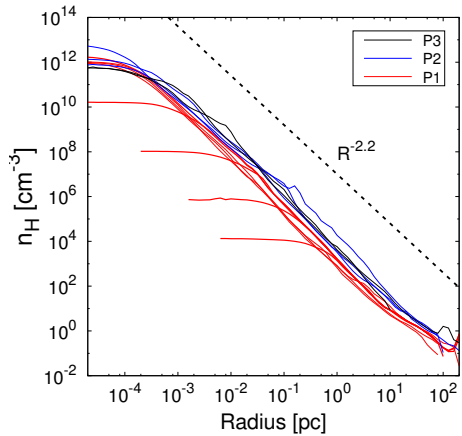}}
\caption{
Radial gas density profiles for 9 selected star-forming clouds. 
The colors indicate the different stellar evolutionary paths
P1 (red, 4 cases), P2 (blue, 3 cases), and P3 (black, 2 cases). 
For one P1 case we also plot the time evolution of the density
profile at the time when the central density is
$10^4$, $10^6$, $10^8$, $10^{10}$, and 
$10^{12} \ {\rm cm^{-3}}$.
For all other clouds, we use snapshots at 
$n_{\rm H,cen} \simeq 10^{12} \ {\rm cm^{-3}}$.
The black dotted line shows
a power-law density distribution with
$n_{\rm H} \propto R^{-2.2}$ for reference.
}
\label{plot_PROFILE_Radi-Dens}
\end{center}
\end{figure}

\subsection{Evolution in the Early Collapse Stage}

\subsubsection{Run-away Collapse of the Clouds}

In this section we describe the early evolution of the star-forming clouds 
up to the moment when a central hydrostatic core is formed by considering
the fate of nine representative cases.
Figure \ref{plot_PROFILE_Radi-Dens} shows that the gravitational collapse of 
a primordial cloud proceeds in the well-known self-similar manner.
The cloud has a central collapsing core and a surrounding envelope during the collapse. 
Whereas the collapsing core has an approximately homogeneous density distribution, 
the envelope develops a power-law profile, $n_{\rm H} \propto R^{-2.2}$ 
\citep[e.g.,][]{omukai98, ripamonti02}. 
Figure \ref{plot_PROFILE_Radi-Dens} also shows the radially-averaged density profiles 
in the nine different clouds at the time when the central density reaches $10^{12} \ {\rm cm^{-3}}$.
We see that densities at the same radial distance can differ 
among the clouds by more than a factor of ten. 
The variation of the density structure is attributed to the
different thermal evolution during the collapse (see Sec. 3.2.2 below). 
Some bumps in the density profiles indicate the presence of neighboring density peaks, 
large disk- or bar-like structure, and/or fragmenting clumps in the collapsing clouds.
We discuss these cases further in Sec. 5.2.2.

\begin{figure}
\begin{center}
\resizebox{7cm}{!}{\includegraphics[clip]{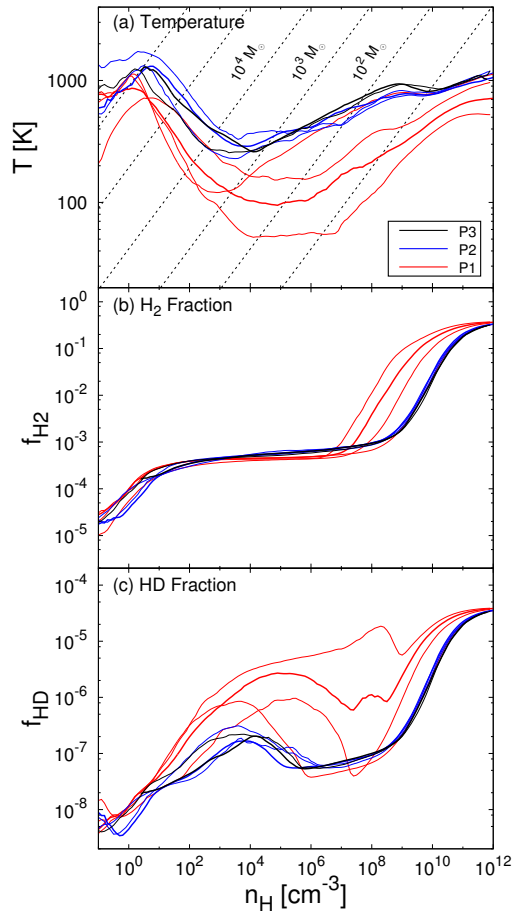}}
\caption{
Gas temperature ($top$ panel), 
${\rm H_2}$ fraction ($middle$), and 
${\rm HD}$ fraction ($bottom$) distributions
as a function of gas number density, $n_{\rm H}$, 
for the same 9 clouds in Fig. \ref{plot_PROFILE_Radi-Dens} 
when the central (highest) density is  $n_{\rm H,cen} = 10^{12} \ {\rm cm^{-3}}$.
We calculate the mass-weighted mean values 
of the gas SPH particles for each density bin.
The dashed lines show the $n_{\rm H}$~-~$T$ relation
for the given Jeans masses.
}
\label{plot_PROFILE_temp_multi}
\end{center}
\end{figure}

\subsubsection{Thermal Evolution During the Collapse}

Next, we address the thermal evolution of the collapsing primordial star-forming clouds. 
We show the distributions of the temperature and density in the envelope 
when the central density is $\simeq 10^{12} \ {\rm cm^{-3}}$
(Figures \ref{plot_PROFILE_Radi-Dens} and \ref{plot_PROFILE_temp_multi}).
The $n$ - $T$ profiles in the envelope at a particular time essentially
trace the thermal evolution of the contracting core
due to the self-similar nature of the collapse.

The thermal evolution of a collapsing cloud is
largely determined by non-equilibrium cooling and heating processes.
The most important coolant of the primordial gas is molecular hydrogen ${\rm H_2}$.
Figures \ref{plot_PROFILE_temp_multi}(a) and (b) show that 
the cloud is heated up to $\sim 1000\ {\rm K}$ due to virialization 
at $n_{\rm H} < 10^2 \ {\rm cm^{-3}}$, 
as the ${\rm H_2}$ fraction slowly increases via the two-body reaction,
\begin{eqnarray}
{\rm H} + {\rm e^-} & \to & {\rm H^-} + h\nu \ , 
\label{eq:H2_two-body1}
\end{eqnarray}
\begin{eqnarray}
{\rm H^-} + {\rm H} &\to & {\rm H_2} + {\rm e^-} \ .
\label{eq:H2_two-body2}
\end{eqnarray}
When a sufficient amount of ${\rm H_2}$ is formed,
the gas temperature begins to decrease to $100 - 200 \ {\rm K}$ 
due to ${\rm H_2}$ line cooling.

Figure \ref{plot_PROFILE_temp_multi}(a) shows that, 
in several cases, the gas temperature becomes lower than 200~K, 
which is unexpectedly low for the so-called Pop III.1 star formation scenario \citep{bromm09}.
The low temperature tracks are more similar to that expected from Pop III.2 star formation 
where HD molecular cooling operates.
Indeed, Figure \ref{plot_PROFILE_temp_multi}(c) shows that 
HD molecules are efficiently formed in these cases, 
suggesting that the low temperatures are due to HD line cooling. 
HD cooling is usually negligible for Pop III.1 star formation
\citep{ripamonti07}.
In the Pop III.2 star formation scenario
${\rm H_2}$ formation is enhanced if the initial electron abundance is high
(see Eqs. \ref{eq:H2_two-body1} and \ref{eq:H2_two-body2}). 
As a result, the temperature becomes low enough to form HD molecules, 
and even drops below $100 \ {\rm K}$ with the additional HD molecular cooling. 
This is not the explanation in our cases, however, since
we do not see a significant enhancement 
of the electron abundance. 

Interestingly, the timescale of cloud collapse 
plays an important role in determining the thermal evolution. 
Although the collapse timescale is always comparable to 
the free-fall timescale for an average given density, 
the collapse can be decelerated, 
e.g., by rotational support if the cloud has finite angular momentum. 
When the collapse is slower, 
more ${\rm H_2}$ molecules are produced 
via reactions (\ref{eq:H2_two-body1}) and (\ref{eq:H2_two-body2}). 
With the enhanced radiative cooling from the more abundant ${\rm H_2}$, 
the temperature becomes sufficiently low to allow HD formation and cooling. 
Note that the slight difference in molecular binding energy 
triggers chemical fractionation at low temperatures.  
Once HD molecular line cooling begins to operate, 
the temperature decreases further down to a few tens of Kelvin. 
In one extreme case, the temperature attains 
the floor of the cosmic microwave background (CMB), 
$T_{\rm CMB} = 2.73 \ (1+z) \simeq 50 \ {\rm K}$ at $z \simeq 20$. 
Thus, we conclude that the effects of a slow collapse are indeed significant.
When the collapse time is only a few times longer than the free-fall time, 
the resulting thermal evolution becomes similar to that of the Pop III.2 case
driven by HD cooling. 
We have also confirmed this behavior by using one-zone models 
for following the thermal evolution during the cloud collapse \citep[e.g.,][]{omukai00}. 
Our detailed analysis is described in Appendix C.

HD cooling becomes inefficient 
at densities greater than $10^8 \ {\rm cm^{-3}}$. 
However, the abundance of ${\rm H_2}$ molecules begins to increase 
as the so-called three-body formation process,
\begin{eqnarray}
{\rm H} + {\rm H} + {\rm H} \to {\rm H_2} + {\rm H} \  ,
\label{eq:H2_three-body}
\end{eqnarray}
comes into play, which converts 
nearly all the hydrogen to molecules 
before the density reaches $10^{12} \ {\rm cm^{-3}}$. 
After that, the cooling process in the collapsing cloud is 
dominated by ${\rm H_2}$ molecules. 

\begin{figure}
\begin{center}
\resizebox{6cm}{!}{\includegraphics[clip]{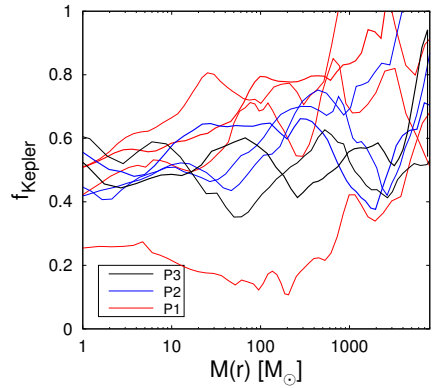}}
\caption{
Averaged profiles of the degree of rotational support, 
$f_{\rm Kepler}$ = $v_{\rm rot} / v_{\rm Kepler}$,
for the same 9 clouds in Fig. \ref{plot_PROFILE_Radi-Dens}
at the moment when $n_{\rm H,cen} = 10^{12} \ {\rm cm^{-3}}$. 
The lowest line represents the case 
for which the cloud is most slowly rotating in our sample
(see Figure \ref{morphology:extremely}a and Table \ref{table:abnormal}).
Generally, $f_{\rm Kepler}$ is large for P1, P2, and P3 in this order.
}
\label{plot_PROFILE_Menc-Fkep}
\end{center}
\end{figure}

\begin{figure}
\begin{center}
\resizebox{7.5cm}{!}{\includegraphics[clip]{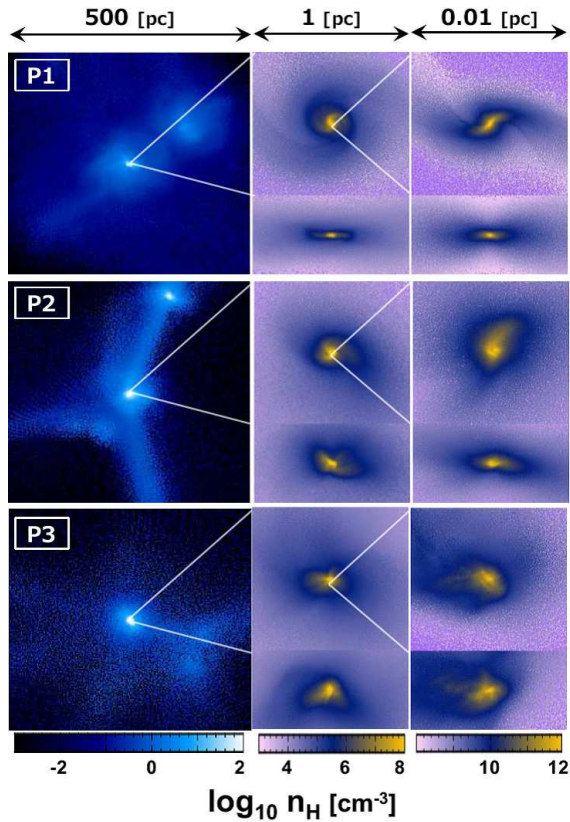}}
\caption{
Projected gas density distributions for three representative cases 
at the moment when $n_{\rm H,cen} = 10^{12} \ {\rm cm^{-3}}$
(corresponding to those denoted 
by thick lines in Figure \ref{plot_PROFILE_Radi-Dens}),
whereby the protostars go through 
three characteristic evolutionary paths (P1, P2, and P3) 
in the later accretion phase. 
The panels show, from left to right, the gas distribution in
regions of 500, 1, and 0.01 pc on a side, respectively.
We show both the face-on and edge-on views 
for the 1 pc and 0.01 pc boxes. 
Note the rotationally supported disk-like structure for the P1 case (top).
}
\label{morphology:classification}
\end{center}
\end{figure}

\subsubsection{Collapse of the Clouds and Angular Momentum}

The angular momentum of a cloud is one of the key factors 
which influences the radial infall velocity and thermal evolution during the collapse.
A cloud with a low angular momentum gravitationally contracts roughly over a
free-fall time. 
Figure \ref{plot_PROFILE_Menc-Fkep} shows for our nine representative cases
the radial distributions of $f_{\rm Kepler}$, the
ratio of the azimuthal rotation velocity $v_{\rm rot}$ 
(defined as the azimuthally averaged velocity perpendicular 
to the total angular momentum vector inside $M(r)$) 
to the Keplerian velocity $v_{\rm Kepler} = \sqrt{GM(r)/r}$.

\begin{figure}
\begin{center}
\resizebox{7cm}{!}{\includegraphics[clip]{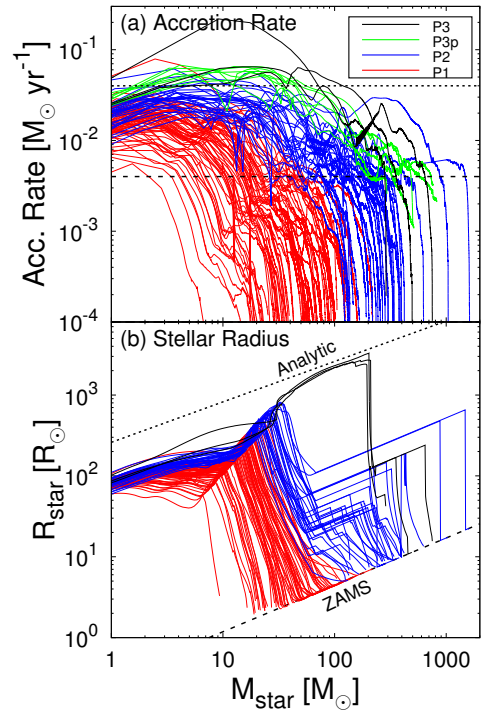}}
\caption{
Accretion rates ($upper$ panel) and 
stellar radii ($lower$ panel) as a function of stellar mass for our entire sample. 
The colors represent the different paths of the protostellar evolution 
as in Fig. \ref{plot_PROFILE_Radi-Dens}. %\uplus 
The two horizontal lines in the $upper$ panel are critical values of accretion rates 
$\dot{M}_{\rm P2} =$ 0.004 ${\rm M_{\odot} \ yr^{-1}}$ (dotted; Eq. \ref{eq:P2-criterion}) and 
$\dot{M}_{\rm P3} =$ 0.04 ${\rm M_{\odot} \ yr^{-1}}$ (dashed; Eq. \ref{eq:P3-criterion}) .
The two diagonal lines in the $lower$ panel indicate
$260 \times M^{0.5} \ {\rm M_{\odot} \ yr^{-1}}$ 
(dotted; super-giant protostar, Eq. \ref{eq:SGPS-radius}) and
$0.31 \times M^{0.58} \ {\rm M_{\odot} \ yr^{-1}}$ 
(dashed; ZAMS track derived from the calculation). 
The green lines in the $top$ panel show the 
results of RHD simulations without the stellar feedback for P3p cases 
(see also Figure \ref{plot_IMF-PopIII_3PATHp}) 
which accurately follow the evolution 
when the UV radiative feedback is ineffective ($\dot{M} < \dot{M}_{\rm P2}$). 
}
\label{plot_EXTEL_multi}
\end{center}
\end{figure}

We see that the profiles of $f_{\rm Kepler}$ are moderately correlated to 
the thermal evolutions of the clouds, 
albeit with a large variance at each mass scale $M(r)$.
On average, clouds with low temperatures  
have higher $f_{\rm Kepler}$  
(recall that the red, blue, and black lines represent clouds with progressively 
lower temperatures in this order; see also Figure \ref{plot_PROFILE_temp_multi}a). 
Interestingly, Figure \ref{plot_PROFILE_Menc-Fkep} also shows that 
the variance of $f_{\rm Kepler}$ decreases with decreasing enclosed mass $M(r)$
whereas the median value is nearly constant around $\simeq 0.5$ \citep[see also][]{abel02}, 
independent of $M(r)$.

The cloud's degree of rotation also influences 
the density distribution during the collapse. 
Figure \ref{morphology:classification} shows 
the projected density distributions for 3 clouds 
with different initial angular momenta.
The top panel clearly shows 
a disk-like structure with notable spiral arms, 
suggesting that a self-gravitating disk is forming around the 
protostar. 
Although the disk-like structure disappears for the lower angular momentum cases, 
the middle and bottom panels still show 
a variety of non-spherical structures.
This suggests a wide range of dynamics in the gravitational collapse of 
the primordial clouds. We will revisit this issue in Section 5.1.

\begin{figure}
\begin{center}
\resizebox{7cm}{!}{\includegraphics[clip]{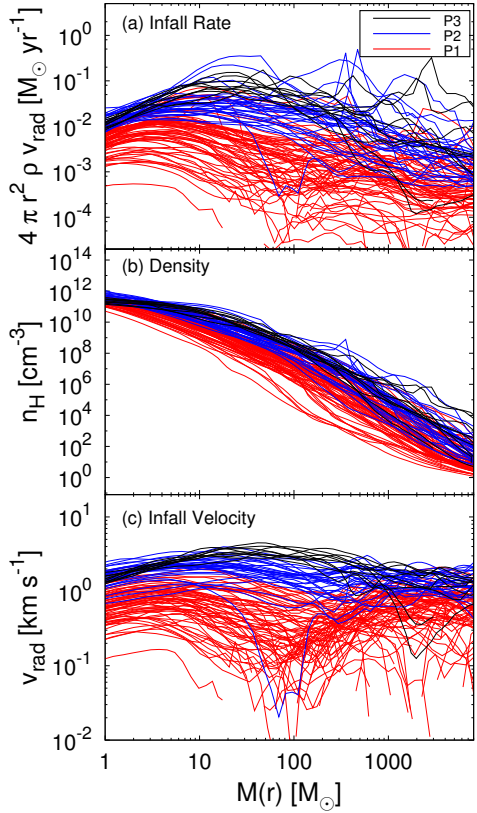}}
\caption{
The averaged radial distribution of the gas infall rate
$4 \pi r^2 \rho(r) v_{\rm rad}(r)$ ($top$ panel), 
the density ($middle$) and infall velocity ($bottom$) 
profiles when $n_{\rm H,cen} = 10^{12} \ {\rm cm^{-3}}$.
The colors represent the three paths of the protostellar evolution 
as in Fig. \ref{plot_PROFILE_Radi-Dens}. 
}
\label{plot_PROFILE_dmdt_multi}
\end{center}
\end{figure}

\subsection{Evolution during the Late Accretion Stage}

\subsubsection{Mass Accretion Histories and the Final Stellar Masses}

We define the late accretion stage as the period during which a
protostar gains most of its final mass. 
Figure \ref{plot_EXTEL_multi}(a) shows the accretion histories 
for the entire sample of 110 protostars. 
The accretion rates gradually decrease with increasing protostellar mass
in all the cases. There are, however, substantial variations among the clouds. 
The accretion rates differ more than a factor of 10 when 
the stellar masses are only a few solar masses.
The protostars' evolution also differs significantly, 
reflecting the variation of the mass accretion rates (Figure \ref{plot_EXTEL_multi}b). 

The stellar effective temperature and the UV luminosity 
rapidly increase when a protostar approaches the ZAMS.
At this point UV feedback becomes efficient and ultimately terminates 
the mass accretion onto the protostar.
For high accretion rates, the protostar reaches the ZAMS when the mass
is typically larger than one hundred solar masses. 
In general, the final stellar masses are greater 
in cases with higher mass accretion rates. 

The wide variety of mass accretion profiles are caused by
differences in the structure of the gas envelope at the birth of the protostellar core. 
Figure \ref{plot_PROFILE_dmdt_multi} shows 
the radially averaged profiles of the mass accretion rate
estimated from the instantaneous density and velocity 
distributions of the envelope, $4 \pi r^2 \rho(r) v_{\rm rad}(r)$.
Note that this is not the accretion history plotted in Fig. \ref{plot_EXTEL_multi}(a).
The ``predicted'' accretion rates differ quite substantially, 
being qualitatively consistent with the actual accretion histories 
found in the radiation hydrodynamic simulations 
(compare Figs. \ref{plot_EXTEL_multi}a and \ref{plot_PROFILE_dmdt_multi}a). 
We find that the thermal evolution during the pre-stellar collapse phase
plays a key role in shaping the structure of the accreting envelope.
Because the thermal evolution depends on the contraction timescale, 
mechanisms that affect the gravitational collapse 
such as the cloud's angular momentum likely affect the overall gas accretion rate.
We discuss the origin of the various accretion histories in Section 4.2.

\begin{figure}
\begin{center}
\resizebox{7cm}{!}{\includegraphics[clip]{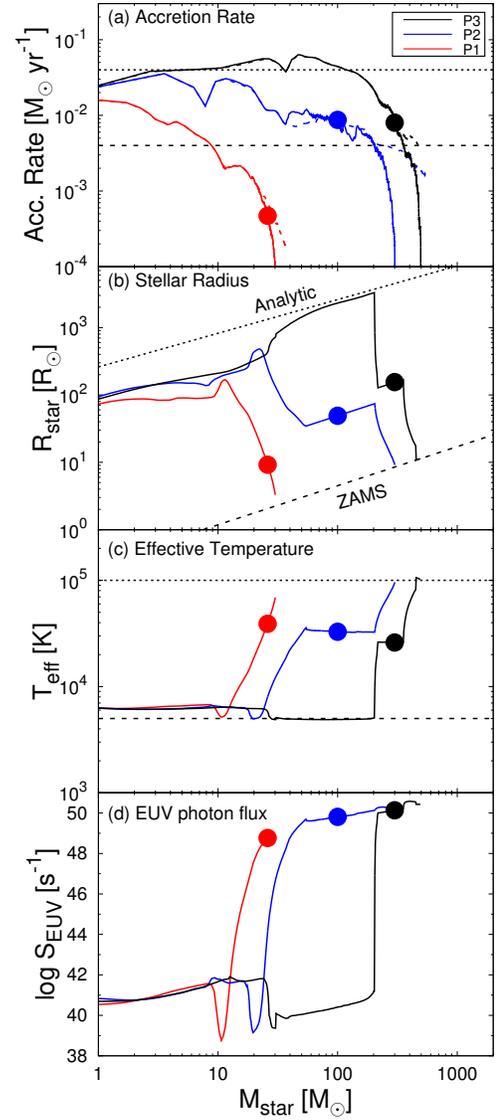}}
\caption{
The accretion rates (top panel: a) and 
stellar properties (b, c, and d) 
versus the stellar mass for the three representative cases 
of the protostellar evolution. 
In panel (a), the solid and dashed lines represent the evolution with and without 
UV radiative feedback from the primordial protostar. 
The dashed and dotted lines in the panels (a) and (b) are 
the same as in Fig. \ref{plot_EXTEL_multi}. 
Two horizontal lines in panel (c) indicate 
$T_{\rm eff} = 10^5 \ {\rm K}$ (dotted) and $5000 \ {\rm K}$ (dashed).
The filled circle marks the epoch
when a bipolar H{\sc ii} region first appears in each case. 
}
\label{plot_EXTEL_Mstr-dMdt-wowi}
\end{center}
\end{figure}

\subsubsection{Three Evolutionary Paths of Accreting Protostars}

Here we focus on three representative cases characterized
by the evolution of the protostellar radius (see Sec. 2.2.1): 
the KH contracting protostar (P1), the oscillating protostar (P2), 
and the super-giant protostar (P3). 
Figure \ref{plot_EXTEL_Mstr-dMdt-wowi} shows
the accretion histories for these three selected cases.
The accretion rates are highest for P3 and lowest for P1 (see also Sec. 2.2.1). 
The figure also presents the accretion histories for the same cases 
but with the UV radiative transfer module switched off,
which demonstrates that stellar UV feedback does indeed
reduce the accretion rates and the final mass of the protostar
late in the evolution.

Case P1 exemplifies the KH contracting protostar 
studied in \cite{hosokawa11, hosokawa12a}. 
The panels (b) - (d) show that, 
in the adiabatic accretion phase when $M_{\rm star} \lesssim 10 \ {\rm M_{\odot}}$, 
the effective temperature and ionizing photon (EUV) luminosity is so low that 
stellar UV feedback is negligible. 
However, the EUV luminosity rapidly increases once the protostar begins KH contraction
at $M_{\rm star} \simeq 10 \ {\rm M_{\odot}}$.
As the stellar radius decreases the effective temperature and the EUV luminosity rise. 
At first, the high accretion rate squelches the H{\sc ii} region
\citep[see e.g.,][]{yorke86, omukai02}, but eventually the
EUV flux is sufficient to ionize the infalling neutral material, driving
an expanding bipolar H{\sc ii} region
into the accretion envelope (also see Fig. \ref{f2}). 
The expanding H{\sc ii} region 
accelerates the envelope gas outward. At the same time
the circumstellar disk is irradiated by the stellar UV,
creating a thermally driven disk wind as the disk gradually photo-evaporates. 
As the accretion rate rapidly decreases,
the protostar contracts faster, further increasing the effective surface temperature,
and the UV feedback becomes stronger.
Figure \ref{plot_EXTEL_Mstr-dMdt-wowi} shows that 
the final stellar mass for the P1 case is fixed at $\simeq 30 \ {\rm M_{\odot}}$, 
just before the star's arrival to the ZAMS.
 
The evolution of P2 looks similar to P1's before KH contraction 
ceases at $M_{\rm star} \simeq 60 \ {\rm M_{\odot}}$, 
when the stellar total luminosity reaches the Eddington value. 
The evolution after this differs from P1's. 
The stellar radius begins to increase with increasing mass,
entering an oscillating phase\footnote{By using our simplified model 
for P2, the radius does {\it not} oscillate, see Appendix B.2.}. 
The effective temperature remains almost constant 
at $\simeq 3 \times 10^4 \ {\rm K}$ during this phase, 
and thus the stellar EUV luminosity does not increase. 
A bipolar H{\sc ii} region first appears late, when 
the stellar mass reaches $\simeq 100 \ {\rm M_{\odot}}$.
The oscillating phase is over at $M_{\rm star} \simeq 200 \ {\rm M_{\odot}}$, 
when the accretion rate falls below the critical rate 
$\dot{M}_{\rm P2} = 4 \times 10^{-3} \ {\rm M_{\odot} \ yr^{-1}}$ 
(Eq. \ref{eq:P2-criterion}), allowing the protostar to resume contraction 
toward the ZAMS. 
Mass accretion onto the star is finally halted
at $M_{\rm star} \simeq 300 \ {\rm M_{\odot}}$. 
It is remarkable that, in this case, a large portion
of the surrounding gas, $200 \ {\rm M_{\odot}}$,
is accreted onto the star well after the breakout of the H{\sc ii} region.

The protostar of our representative P3 case enters the super-giant phase
soon after the adiabatic accretion phase, when
$M_{\rm star} \simeq 30 \ {\rm M_{\odot}}$. 
The stellar radius monotonically increases 
with increasing mass, according to the relation (\ref{eq:SGPS-radius}). 
The radius exceeds $3000 \ {\rm R_{\odot}}$ 
when the stellar mass is $M_{\rm star} \simeq 200 \ {\rm M_{\odot}}$. 
The effective temperature is still around 5000 K during this super-giant
phase, which means that the EUV luminosity is too low to launch an H{\sc ii} region. 
The protostar begins to contract after the accretion rate falls below 
a few $\times 10^{-2} \ {\rm M_{\odot} \ yr^{-1}}$, 
the critical rate given by Eq. (\ref{eq:P3-criterion}). 
The final contraction is very rapid 
because the KH timescale of the super-giant protostar is very short, 
less than $10^3$ years \cite[e.g.,][]{hosokawa12a}. 
Shortly thereafter, KH contraction is interrupted, and
the protostellar evolution becomes similar to P2.
The H{\sc ii} region begins expanding
when the stellar mass is $\simeq 300 \ {\rm M_{\odot}}$, which in turn
reduces the accretion rate below the critical value $\dot{M}_{\rm P2}$. 
As the star contracts once again toward the ZAMS, 
the stellar UV feedback becomes stronger, 
and the stellar mass is finally fixed 
at $M_{\rm star} \simeq 500 \ {\rm M_{\odot}}$. 

We note that, in the cases of P2 and P3, the effective temperature of the expanding
protostar remains low, which mostly explains 
why the breakout of the H{\sc ii} region occurs late, 
when the mass is $\simeq 200$ and $300 \ {\rm M_{\odot}}$, respectively. 
However, Figure \ref{plot_EXTEL_Mstr-dMdt-wowi}(d) also shows that 
the EUV photon luminosity is in fact higher than the case of P1 
when the H{\sc ii} region first emerges. 
This is explained by the fact that the accretion envelope is denser 
in P2 and P3 because of the higher accretion rates. 
More ionizing photons are needed to create an H{\sc ii} in dense regions, 
where hydrogen atoms recombine rapidly \citep[also see][]{hosokawa12b}.

\section{Origin of the Wide Range of Stellar Masses}

Our simulations have followed the process of primordial star formation 
in a number of cosmological halos until the protostars formed reach the ZAMS.
We here examine the origin of the wide distribution of stellar masses. 
First of all, Figure \ref{plot_EXTEL_multi} shows a
clear correlation between the accretion rates and the final stellar masses.
We have already seen that 
the accretion rate onto a protostar is largely determined by
the density and velocity structure of 
the envelope around the protostar (Fig. \ref{plot_PROFILE_dmdt_multi}). 
We now study in greater detail the formation process of the star-forming gas clouds, 
whose initial conditions are set at the virialization of the host dark halos.

We shall focus on the overall gas infall rates at two different 
mass scales; the parent gas cloud and the host dark matter halo.
We show that the final stellar masses
are indeed correlated with the gas infall rates (Sec. 4.1).
We then study how the infall rates are related to 
other properties of the gas clouds and those of the host halos (Sec. 4.2).

\begin{figure}
\begin{center}
\resizebox{7cm}{!}{\includegraphics[clip]{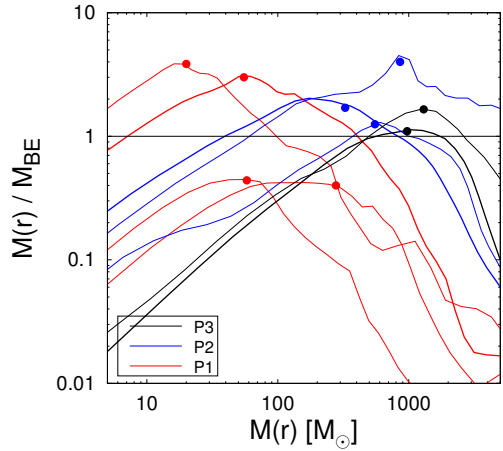}}
\caption{
The ratio of the enclosed gas mass to the local BE mass for 9 samples 
when the extremum of $M_{\rm enc}(r) / M_{\rm BE}$ reaches the maximum during the collapse. 
Filled circles mark the positions where $M(r)/M_{\rm BE}$ takes the maximum, 
by which we define the cloud mass.
}
\label{plot_100FS_Menc-MeMbe}
\end{center}
\end{figure}

\subsection{Stellar Masses and Gas Infall Rates}

\subsubsection{Infall Rates in Gas Clouds}

A characteristic mass scale of a gravitationally contracting gas cloud 
would be the Bonner-Ebert (BE) mass \citep{bonnor56, ebert55}
\begin{eqnarray}
M_{\rm BE} & = & \frac{1.18 c_{\rm s}^4}{G^{\rm 3/2} P_{\rm ext}^{1/2}} \ {\rm M_{\odot}} \ , \\
& \approx & 1050 \ {\rm M_{\odot}} \cdot \left( \frac{T}{200 \ {\rm K}} \right)^{3/2} \left( \frac{\mu}{1.22} \right)^{-2} \nonumber \\ 
& & \cdot \left( \frac{n_{\rm H}}{10^4 \ {\rm cm^{-3}}} \right)^{-1/2}  \left( \frac{\gamma}{1.66} \right)^2 \ ,
\label{eq:M_BE}
\end{eqnarray}
where 
$c_{\rm s}$ is the sound speed,
$P_{\rm ext}$ is the external pressure,
$n_{\rm H}$ is the hydrogen number density, 
$\mu$ is the mean molecular weight, and
$\gamma$ is the adiabatic index.
A cloud is expected to collapse  
when the mass within a given radius exceeds the local BE mass, 
$M_{\rm enc}(r) > M_{\rm BE}$.
For a primordial gas cloud, this run-away collapse occurs when 
the first temperature dip appears in the thermal evolutionary track 
(Fig. \ref{plot_PROFILE_temp_multi}a), 
i.e., at the so-called loitering regime \citep[e.g.,][]{bromm02}.
In several of our dark matter halos
run-away collapse occurs even though the ratio 
$M_{\rm enc}(r) / M_{\rm BE}$ is (slightly) smaller than unity 
(Fig. \ref{plot_100FS_Menc-MeMbe}). 
This is because the above BE mass provides a critical mass for 
gravitational stability only under some idealized conditions, 
e.g., assuming static motion for the initial condition, 
which are not realized in our 3D simulations, and the isothermal equation of state.  
Here, we shall define the mass of the collapsing cloud $M_{\rm cloud}$ to be 
the mass within a radius at which the ratio of the enclosed mass 
to the BE mass takes the maximum value during the collapse
(see Figure \ref{plot_100FS_Menc-MeMbe}). 
Typically, a primordial gas cloud that cools via ${\rm H_2}$ line cooling has 
a large mass $M_{\rm cloud} \sim 1000 \ {\rm M_{\odot}}$. 
Note that HD line cooling also operates in some clouds. 
Such clouds have relatively small masses $M_{\rm cloud} \sim$ 10 - 100 ${\rm M_{\odot}}$
when they collapse gravitationally (Section 3.2.2). 

\begin{figure*}
\begin{center}
\resizebox{15cm}{!}{\includegraphics[clip]{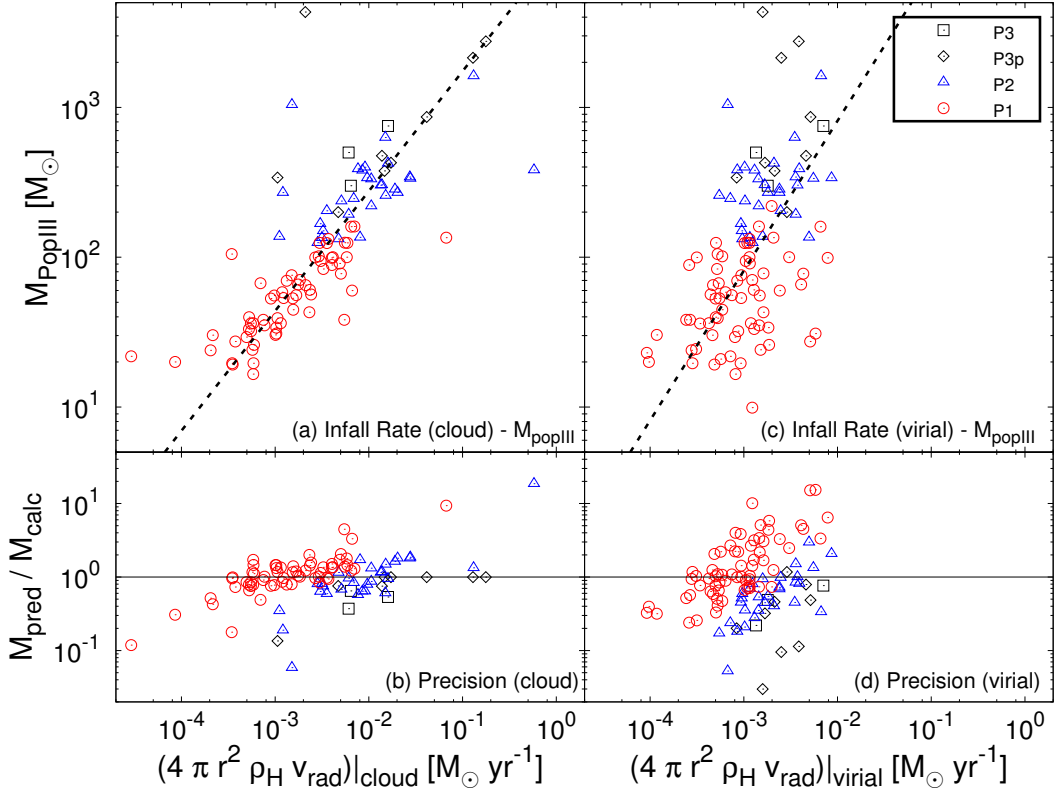}}
\caption{
Final stellar masses as a function of gas infall rates over
cloud scales ($left$) and over dark matter halo scales ($right$).
Three colors of symbols represent the evolutionary 
paths of the protostellar evolution 
as in Figure \ref{plot_IMF-PopIII_3PATHp}. 
The dashed lines represent the fitting functions 
given by Eqs. (\ref{eq:MpopIII-dMdtJeans}) and (\ref{eq:MpopIII-dMdtVirial}) . 
In the bottom frames we plot the ratios of the actual 
stellar masses to the analytic fitting functions.
}
\label{plot_LIST_dmdt_multi2}
\end{center}
\end{figure*}

In Sec. 3.2, we have seen that the stars formed are more massive
for more rapid accretion (Fig. \ref{plot_EXTEL_multi}), and that 
the overall accretion rate can be estimated approximately from the density and 
velocity structure of the gas envelope at the time when a hydrostatic 
protostellar core is formed
(Fig. \ref{plot_PROFILE_dmdt_multi}). 
Figure \ref{plot_LIST_dmdt_multi2}(a) supports this notion, 
showing a good correlation between the final stellar masses and 
the accretion rates estimated from $\dot{M}_{\rm cloud} = 4 \pi r^2 \rho(r) v_{\rm rad}(r)$ 
at the cloud mass scale. We evaluate these quantities at radius $r_{\rm cloud}$
within which the enclosed gas mass is $M_{\rm cloud}$.

Our fitting formula for the correlation
\begin{equation}
M_{\rm popIII} = 100 \ {\rm M_{\odot}} \left( \frac{\dot{M}_{\rm cloud}}{2.8 \times 10^{-3} \ {\rm M_{\odot} \ yr^{-1}}} \right)^{0.8} \ ,
\label{eq:MpopIII-dMdtJeans}
\end{equation}
is shown in Figure \ref{plot_LIST_dmdt_multi2}(a). 
Figure \ref{plot_LIST_dmdt_multi2}(b) shows that the the deviations from the 
above relation are indeed small. 
This relationship should be quite useful, because it allows us to 
estimate the final stellar mass
from $\dot{M}_{\rm cloud}$ without following the detailed
protostellar evolution.

\begin{table*}[t]
\caption{Averaged Properties for Three Evolution Paths\label{table:average}}
\begin{center}
\begin{tabular}{crccrc}
\tableline\tableline\noalign{\smallskip}
Path & $\overline{M}_{\rm popIII}$ & $\overline{z}_{\rm form}$ & $\overline{M}_{\rm virial}$ & $\overline{M}_{\rm cloud}$ & $\overline{\beta}_{\rm coll}$ \\
 & $({\rm M_{\odot}})$ & & $(10^5 \ {\rm M_{\odot}})$ & $({\rm M_{\odot}})$ & \\
\noalign{\smallskip}\tableline\noalign{\smallskip}
P1 &  51.2 $\pm$ \ \ 4.2 & 18.03 $\pm$ 0.55 & 3.13 $\pm$ 0.20 & 170.5 $\pm$ \ 20.3 & 0.353 $\pm$ 0.017\\
P2 &  282.1 $\pm$ \ 30.3 & 18.86 $\pm$ 0.77 & 4.06 $\pm$ 0.35 & 626.3 $\pm$ \ 80.6 & 0.298 $\pm$ 0.053\\
P3 &  483.0 $\pm$ 128.3 & 17.57 $\pm$ 0.66 & 5.06 $\pm$ 0.55 & 837.9 $\pm$ 118.9 & 0.206 $\pm$ 0.064\\
\noalign{\smallskip}\tableline\noalign{\smallskip}
\end{tabular}
\tablecomments{
All quantities are the averaged amount for each evolutionary path.
Column 2: Final stellar mass, 
Column 3: Redshift when the central density reaches $10^6 \ {\rm cm^{-3}}$, 
Column 4: Mass scale of the virialized dark matter halo, 
Column 5: Mass scale of the star-forming cloud, and
Column 6: The ratio of the rotation to gravitational energy of the cloud.
}
%\tablerefs{}
\end{center}
\end{table*}

%%%
%\begin{deluxetable}{crccrc}
%\tablewidth{0pt}
%\tablenum{3}
%\tablecaption{Averaged Properties for Three Evolution Paths}
%\tablehead{
%\colhead{Path} & 
%\colhead{$\overline{M}_{\rm popIII}$} &
%\colhead{$\overline{z}_{\rm form}$} & 
%\colhead{$\overline{M}_{\rm virial}$} &
%\colhead{$\overline{M}_{\rm cloud}$} &
%\colhead{$\overline{\beta}_{\rm coll}$} \\
%\colhead{} & 
%\colhead{$({\rm M_{\odot}})$} &
%\colhead{} & 
%\colhead{$(10^5 \ {\rm M_{\odot}})$} &
%\colhead{$({\rm M_{\odot}})$} &
%\colhead{}}
%\startdata
%P1 &  51.2 $\pm$ \ \ 4.2 & 18.03 $\pm$ 0.55 & 3.13 $\pm$ 0.20 & 170.5 $\pm$ \ 20.3 & 0.353 $\pm$ 0.017\\
%P2 &  282.1 $\pm$ \ 30.3 & 18.86 $\pm$ 0.77 & 4.06 $\pm$ 0.35 & 626.3 $\pm$ \ 80.6 & 0.298 $\pm$ 0.053\\
%P3 &  483.0 $\pm$ 128.3 & 17.57 $\pm$ 0.66 & 5.06 $\pm$ 0.55 & 837.9 $\pm$ 118.9 & 0.206 $\pm$ 0.064\\
%\enddata
%\tablecomments{
%All quantities are the averaged amount for each evolutionary path.
%Column 2: Final stellar mass, 
%Column 3: Redshift when the central density reaches $10^6 \ {\rm cm^{-3}}$, 
%Column 4: Mass scale of the virialized dark matter halo, 
%Column 5: Mass scale of the star-forming cloud, and
%Column 6: The ratio of the rotation to gravitational energy of the cloud.
%}
%\tablerefs{}
%\label{table:average}
%\end{deluxetable}
%%%

\subsubsection{Infall Rates at Large Scales}

Primordial gas clouds are formed in early dark matter halos,
and one can thus expect that some properties of the host
dark halos will affect the structure of the gas clouds and possibly
the formation of primordial stars within them as well.
Figure \ref{plot_LIST_dmdt_multi2}(c) shows that 
the final stellar masses are indeed correlated 
with the mass infall rates at the halo scale.
We evaluate the density and radial velocity at the virial radius of a halo $r_{\rm virial}$.
Namely, we determine the infall rate from 
$\dot{M}_{\rm virial} = 4 \pi r_{\rm virial}^2 \rho(r_{\rm virial}) v_{\rm rad}(r_{\rm virial})$. 
Similar to Eq. (\ref{eq:MpopIII-dMdtJeans}), we find a good empirical fit
\begin{equation}
M_{\rm popIII} = 100 \ {\rm M_{\odot}} \ \left( \frac{\dot{M}_{\rm virial}}{1.2 \times 10^{-3} \ {\rm M_{\odot} \ yr^{-1}}} \right) \ ,
\label{eq:MpopIII-dMdtVirial}
\end{equation}
which indicates that the mass infall rates at the halo scale are 
roughly proportional to those at the cloud scale. 
The correlation at the halo scale is somewhat weaker than that at the cloud scale.
This is probably because there is a time gap  
between the virialization of the halos and the onset of collapse. 
Nonetheless, Equation (\ref{eq:MpopIII-dMdtVirial}) suggests that 
the chemo-thermal evolution of the star-forming cloud, 
the resulting accretion history and 
the final stellar mass are affected, at least partially,
by the very early conditions of the dark matter halos.

\begin{figure*}
\begin{center}
\resizebox{15cm}{!}{\begin{tabular}{cc}
\includegraphics[clip,scale=0.6]{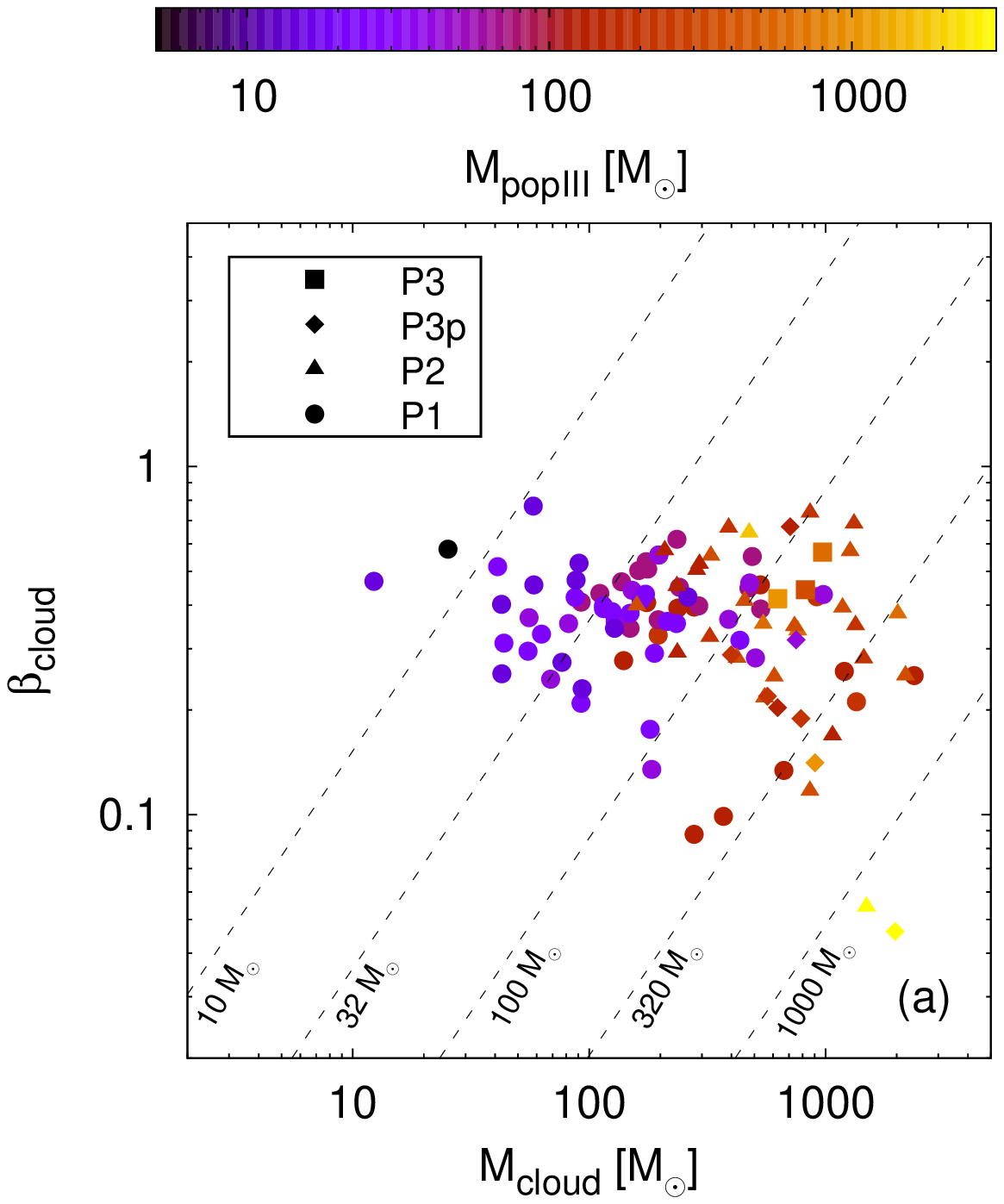} &
\includegraphics[clip,scale=0.6]{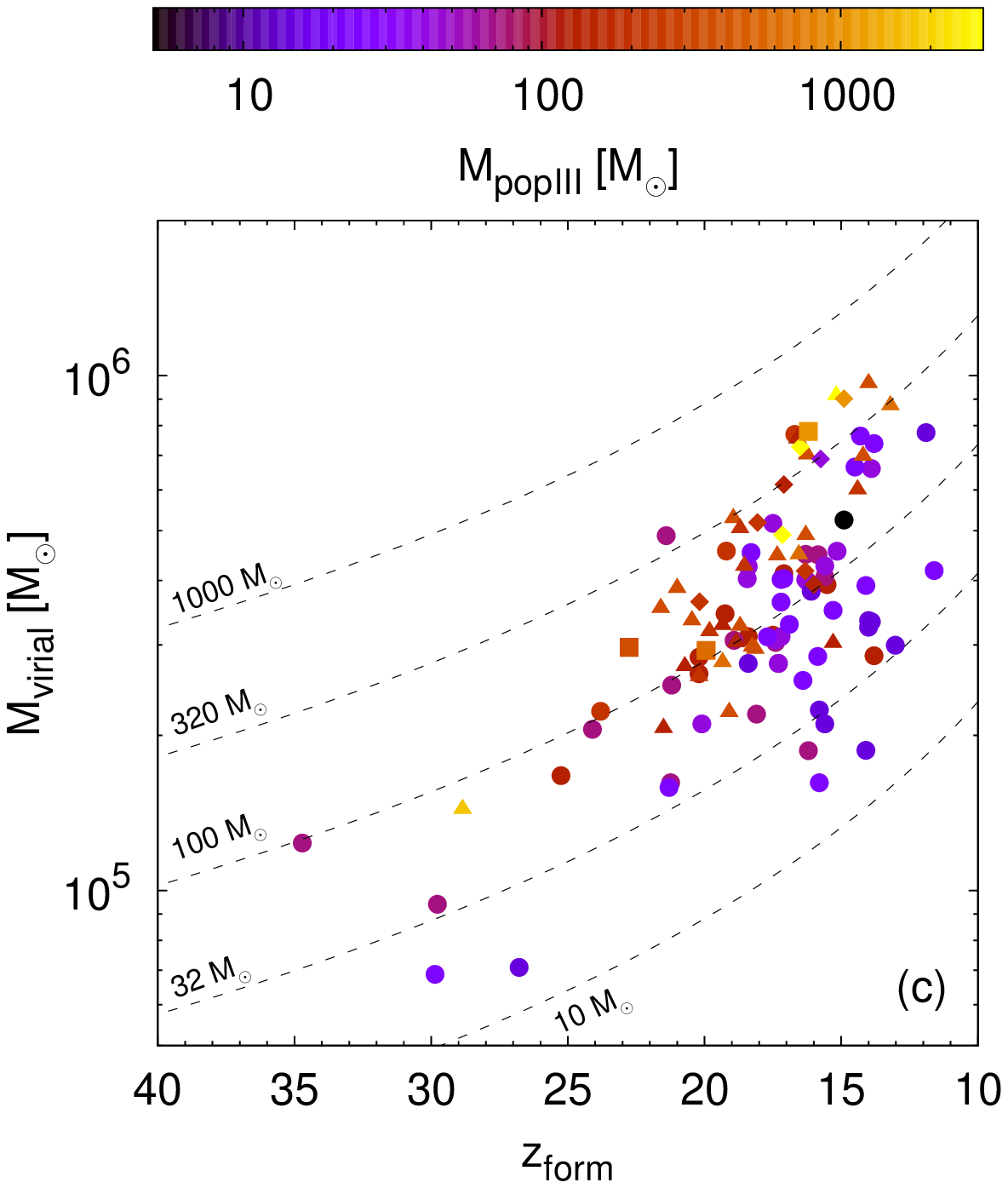} \\
\includegraphics[clip,scale=0.6]{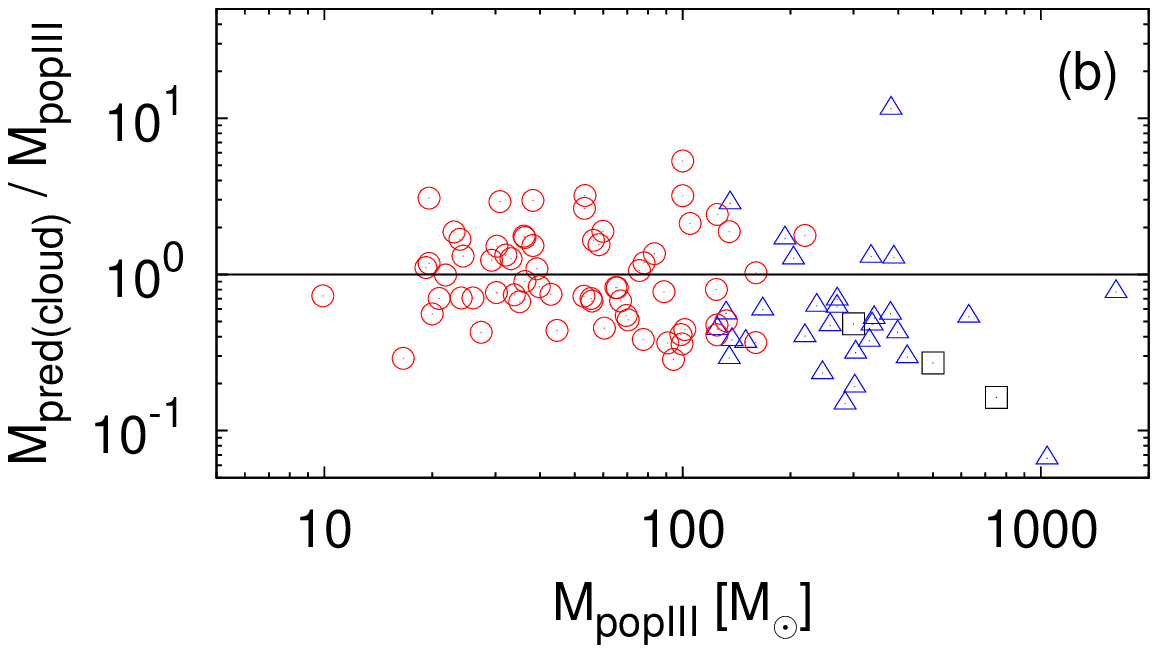} &
\includegraphics[clip,scale=0.6]{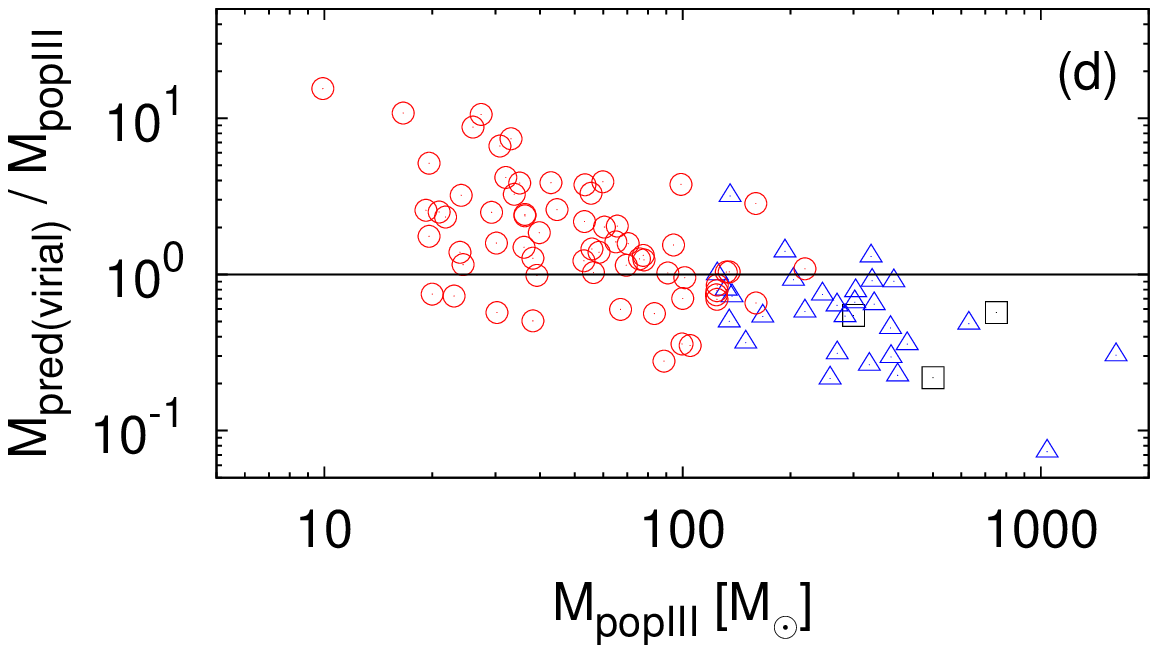}
\end{tabular}}
\caption{
$Top$: The rotation parameter $\beta_{\rm cloud}$ for each of the sample
of 110 cloud versus the cloud mass $M_{\rm cloud}$ (left) 
and the halo mass as a function of the formation redshift $z_{\rm form}$ (right).
The different symbols denote the three paths of the protostellar evolution, 
and the colors are indicative of the final stellar masses (see color scale at the
top). The dashed lines represent the fitting functions given by 
Eqs. (\ref{eq:Mest-Jeans}) and (\ref{eq:Mest-Virial})
for $M_{\rm popIII}$ = 10, 32, 100, 320, and 1000 ${\rm M_{\odot}}$.
$Bottom$: ratios of estimated masses by 
Eqs. (\ref{eq:Mest-Jeans}) and (\ref{eq:Mest-Virial}) to final stellar masses.
}
\label{plot_LIST_Estimation-MpopIII}
\end{center}
\end{figure*}

\subsection{Key Quantities Shaping the Distribution of Final Stellar Masses}

We have found a positive correlation 
between the final stellar masses and the gas infall rates 
(Fig. \ref{plot_LIST_dmdt_multi2}). 
This motivates us to further investigate the origin of 
the diversity of infall rates. 
To this end we first consider the effects of rotation.

\subsubsection{Correlation at the Gas Cloud Scale}

To first order the gas mass accretion rate of a Jeans unstable cloud is 
approximately
\begin{equation}
\dot{M} \sim \frac{M_{\rm Jeans}}{t_{\rm ff}} \propto T_{\rm Jeans}^{1.5} \ ,
\label{eq:infall-rate}
\end{equation}
where 
$M_{\rm Jeans} \propto c_{\rm s}^3/\rho^{1/2}$ is the Jeans mass and
$t_{\rm ff} = \sqrt{(3 \pi)/(32 G \rho)} \propto \rho^{-1/2}$ is the free-fall time.
Note that the collapsing cloud mass is 
not strictly equal to the Jeans mass 
($M_{\rm BE}$; e.g., Figure \ref{plot_100FS_Menc-MeMbe}).
Gravitational collapse does not always proceed 
over a free-fall timescale because rotation can prevent 
or delay the collapse.
The actual {\it collapse time} $t_{\rm cloud}$ 
likely depends on the rotation parameter, 
the ratio of the rotational energy to gravitational energy,
\begin{equation}
\beta_{\rm cloud} = \frac{\Omega_{\rm cloud}^2 R_{\rm cloud}^3}{3 G M_{\rm cloud}} \ ,
\label{eq:beta}
\end{equation}
where $\Omega(r)$ = $| (\vec{r} \times \vec{v (r)})/r^2 | = v_{\perp}(r) / r$
with $v_{\perp} (r)$ at distance $r$ being the velocity perpendicular to the rotational axis. 
We define $\Omega_{\rm cloud}$ at the cloud radius.
Naively, we expect that a cloud with a large $M_{\rm cloud}$ and a low $\beta_{\rm cloud}$ 
collapses quickly and therefore that the gas accretion rate onto the protostar
in the cloud is large (also see Table \ref{table:average}).

Figure \ref{plot_LIST_Estimation-MpopIII}(a) shows
how the final stellar masses depend on the various cloud masses and rotation parameters.
As expected, we see that 
the final stellar mass is higher for larger cloud mass and/or 
for lower rotation parameter. 
We find the dependence is approximately
\begin{eqnarray}
M_{\rm popIII} = 100 \ {\rm M_{\odot}} \left( \frac{M_{\rm cloud}}{{\rm 350 \ M_{\odot}}} \cdot \frac{0.3}{\beta_{\rm cloud}} \right)^{0.8} \ ,
\label{eq:Mest-Jeans}
\end{eqnarray}
which also shows the dependences of the infall rates 
on Eq. (\ref{eq:MpopIII-dMdtJeans}), 
\begin{eqnarray}
\dot{M}_{\rm cloud} = 2.8 \times 10^{-3} \ {\rm M_{\odot} \ yr^{-1}} \left( \frac{M_{\rm cloud}}{{\rm 350 \ M_{\odot}}} \cdot \frac{0.3}{\beta_{\rm cloud}} \right) \ .
\label{eq:dMdt-Jeans}
\end{eqnarray}

The relation (\ref{eq:Mest-Jeans}) is plotted for several different stellar masses
in Figure \ref{plot_LIST_Estimation-MpopIII}(a).  
Whereas Eq. (\ref{eq:Mest-Jeans}) approximately follows the 
variations of the stellar mass, there are substantial deviations from 
the fit (Figure \ref{plot_LIST_Estimation-MpopIII}b).

\subsubsection{Correlation at the Dark Halo Scale}

We also find that the infall rates at
the halo virial radii are correlated with two quantities,
the redshift at which a halo forms $z_{\rm form}$
and the mass of the dark halo $M_{\rm virial}$. 
Thus the resulting final stellar mass likely depends on 
these parameters of the host halo. 
Figure \ref{plot_LIST_Estimation-MpopIII}(c) shows 
the final stellar masses 
for different sets of $z_{\rm form}$ and $M_{\rm virial}$. 
We see that massive stars preferentially form at higher redshifts 
and in more massive dark halos. 
A good fit providing the estimate for the stellar mass is 
\begin{eqnarray}
M_{\rm popIII} = 100 \ {\rm M_{\odot}} \left( \frac{1+z}{20} \right)^{3} \left( \frac{M_{\rm virial}}{{\rm 3 \times 10^5 \ M_{\odot}}} \right)^{2} \ ,
\label{eq:Mest-Virial}
\end{eqnarray}
which with use of Eq. (\ref{eq:MpopIII-dMdtVirial}) can be converted to 
\begin{eqnarray}
\dot{M}_{\rm virial} = 1.2 \times 10^{-3} \ {{\rm M_{\odot} \ yr^{-1}}} \left( \frac{1+z}{20} \right)^{3} \left( \frac{M_{\rm virial}}{{\rm 3 \times 10^5 \ M_{\odot}}} \right)^{2} \ .
\label{eq:dMdt-Virial}
\end{eqnarray}
Although there is substantial scatter
for small cloud masses (Fig. \ref{plot_LIST_Estimation-MpopIII}d), 
Eq. (\ref{eq:Mest-Virial}) provides
a reasonable estimate of the final stellar mass 
using basic properties of the host dark halo. 

It is interesting to study why the two parameters $M_{\rm virial}$ and $z$ control 
the infall rates at the halo virial radius,
and why the final stellar masses correlate with the large-scale infall rates. 
As shown in Sec. 3.2.2 (and in Appendix C), 
the final stellar mass sensitively depends on the thermal evolution of a collapsing cloud. 
Let us suppose that the infall rate is approximately proportional to 
$M_{\rm virial}/t_{\rm ff}$ with some variance due to rotation.
Since the mean density within a collapsed halo scales
with the mean density of the universe at the epoch $z$ as 
\begin{equation}
\rho_{\rm crit}(z) \equiv \frac{3 H^2(z)}{8 \pi G} \ , \frac{H^2(z)}{H^2_0} \simeq \Omega_{\rm m} (1+z)^3 + \Omega_{\rm \Lambda} \ ,
\end{equation}
the above estimate should display a dependence
$M_{\rm virial}/t_{\rm ff} \propto M_{\rm virial} (1+z)^{3/2}$.
This qualitatively explains the overall trend, but
we find that the actual dependence on redshift is stronger (Eq. \ref{eq:dMdt-Virial}).

The additional dependence could be attributed to the degree of halo spin.
Our samples show that on average, at a fixed halo mass, 
the spin parameter decreases with increasing redshift.
At high redshifts, dark matter halos with a given mass are formed 
from high $\sigma$ fluctuations of the initial density field.
Such halos acquire systematically smaller amounts of angular momentum 
via tidal interaction with the surrounding structure.
The rotational support at the halo scale would be thus inefficient 
at the higher redshift, which explains the additional dependence 
on redshift in Eq. (\ref{eq:dMdt-Virial}).

It is worth comparing our findings with the conclusion of previous studies.
\cite{gao07} present simulations of eight primordial gas clouds.
Their gas clouds show significant scatter in the instantaneous mass accretion rate, 
over one order of magnitude (see their Figure 11). 
However, they did not study in detail correlations between the accretion rate 
and halo mass etc., and thus the origin of scatter 
among the collapsing clouds remained unclear.
It is likely that the scatter is caused by a combination of rotation,
halo mass, and the collapse epoch, as we have discussed in this section.
\cite{o'shea07} present a statistical study of primordial cloud formation.
On the assumption that the stellar mass is proportional to 
the overall gas infall rate at some initial epoch, they argue that 
smaller mass stars would form at higher redshifts. 
This appears inconsistent with our results that $M_{\rm popIII} \propto (1+z)^3$.
\cite{o'shea07} claim that 
the redshift-dependence of the virial temperature of a halo with a given mass,
\begin{equation}
T_{\rm virial} \propto M_{\rm virial}^{2/3} \ (1+z) \ ,
\label{eq:Tvirial}
\end{equation}
is essential for determining accretion rates onto the protostar. 
In an early stage of primordial cloud formation, 
${\rm H_2}$ molecules are produced via the ${\rm H^-}$ channel 
(Eqs. \ref{eq:H2_two-body1} and \ref{eq:H2_two-body2}), 
which occurs efficiently at higher temperatures. 
The above dependence of $T_{\rm virial} \propto (1+z)$ can be understood
because slightly more ${\rm H_2}$ molecules are formed 
in a cloud at high redshifts. The resulting temperature of the cloud 
is lower, and the mass accretion rate onto a protostar would be correspondingly lower 
because of the relation $\dot{M} \propto T^{1.5}$.
This argument holds for halos with a fixed mass.
We have a larger number of halos that have a wide range of $M_{\rm virial}$. 
For our cosmological halos that form at different redshifts, 
the virial temperatures do not simply scale with the collapse redshifts
\footnote{The set of samples in \cite{o'shea07} also shows a large scatter of halo mass in the range of 
$1.5 \times 10^5 \ {\rm M_{\odot}}< M_{\rm virial} < 7 \times 10^5 \ {\rm M_{\odot}}$.
However, they suggest that the accretion rates depend only on the collapse redshift. 
We speculate the apparent discrepancy is seen simply because the statistical set of 12 samples 
is too small to find the dependence on two variables, 
$\{z_{\rm virial}, M_{\rm virial}\}$.}.
We actually find an approximately constant 
$T_{\rm virial} \simeq 820 \ {\rm K}$ for collapse. 
This is reasonable because
chemical reactions should take place at the same rate at a constant
gas temperature \citep[e.g.,][]{glover13}. 
Therefore, if the virial temperature (instead of mass) is fixed,
the corresponding halo mass is lower at higher redshifts, 
which cancels out the $z$-dependence in Eq. (\ref{eq:Tvirial}). 

We note that, whereas the equilibrium ${\rm H_2}$ abundance is an increasing function of the temperature, 
the amount of ${\rm H_2}$ molecules necessary to cool the gas is 
a {\it decreasing} function of the gas temperature \citep[see Figure 2 in][]{glover13}. 
The balance between the two processes sets the critical temperature
above which the gas can cool and condense \citep{yoshida03a}. 
Because we select the dark halos where the primordial gas clouds 
are formed via ${\rm H_2}$ cooling, 
the gas temperatures as well as ${\rm H_2}$ abundances do not differ much
between members of our sample. 
Note also that the $z$-dependence in our relation (\ref{eq:Mest-Virial}) 
is attributed to the collapse time 
rather than the virial temperature, as explained above. 
Overall, with our large number of gas clouds, we are able to explore
a large parameter space such as halo mass and collapse epoch
and thus are able to find correlations between multiple quantities. 

\section{Discussion}

\subsection{Nature of Dark Matter Halos}

\begin{figure}
\begin{center}
\resizebox{7.5cm}{!}{\includegraphics[clip]{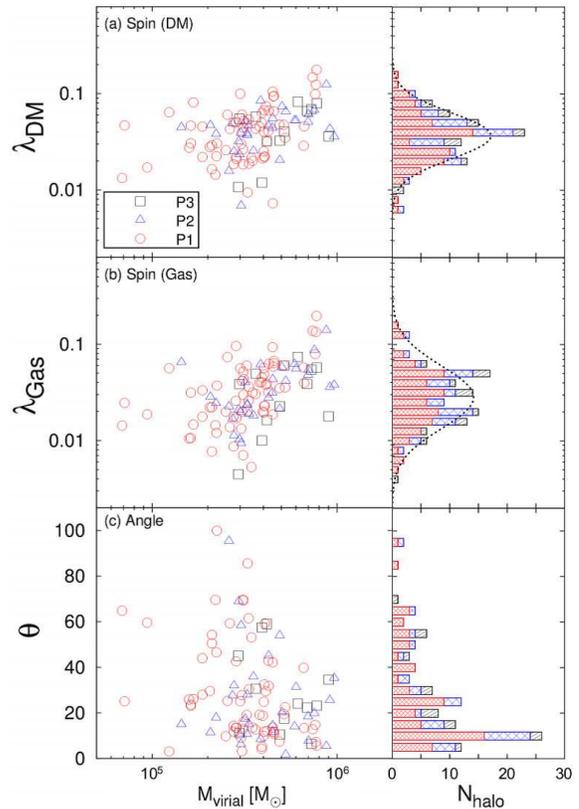}}
\caption{
Distributions of spin parameters for dark matter ($top$ panels) and for gas ($middle$) inside virialized dark matter halos. 
The $bottom$ panel shows the angles between the angular momentum vectors of these two components.
In each panel, the $right$ frame shows the projected histograms summed over all masses.
The dotted lines show best-fit lognormal distributions (Eq. \ref{eq:fit-lognormal}) with
$\overline{\lambda}_{\rm DM} = 0.0495$, $\sigma_{\lambda_{\rm DM}} = 0.545$ for dark matter and
$\overline{\lambda}_{\rm gas} = 0.0498$, $\sigma_{\lambda_{\rm gas}} = 0.750$ for gas.
}
\label{fig-spin}
\end{center}
\end{figure}

The physical properties of the star-forming clouds may depend on 
the initial angular momentum of the host halo.
To study the angular momenta of dark matter 
and gas, we calculate the spin parameter following the 
definition of \cite{bullock01}:
\begin{eqnarray}
\lambda' \equiv \frac{j_{\rm vir}}{\sqrt{2} R_{\rm vir} V_{\rm vir}} \ ,
\label{eq:spin-param}
\end{eqnarray}
where $j_{\rm vir}$ is the specific angular momentum and
$V_{\rm vir} = \sqrt{G M_{\rm vir}/R_{\rm vir}}$ is the circular velocity of the halo.
We also compute the angle between the two angular momentum vectors
\begin{eqnarray}
\theta = \cos^{-1} \left[ \frac{{\bf J}_{\rm DM} \cdot {\bf J}_{\rm gas}}{|{\bf J}_{\rm DM}| |{\bf J}_{\rm gas}|} \right] \ .
\label{eq:spin-angle}
\end{eqnarray}
Figure \ref{fig-spin} shows the range of spin parameters and the alignment angles
present in our sample.

We confirm that the distribution of the spin parameters can be fitted 
by a lognormal distribution (dotted lines in Figure \ref{fig-spin})
\begin{eqnarray}
p(\lambda) d\lambda = \frac{1}{\sqrt{2 \pi} \sigma_{\lambda}} \exp \left[ - \frac{\ln^2 (\lambda / \overline{\lambda})}{2 \sigma_{\lambda}} \right] \frac{d\lambda}{\lambda} \ ,
\label{eq:fit-lognormal}
\end{eqnarray}
with $\overline{\lambda}_{\rm DM} = 0.0495,\ \sigma_{\lambda_{\rm DM}} = 0.545$ for the dark matter 
whereas
$\overline{\lambda}_{\rm gas} = 0.0498,\ \sigma_{\lambda_{\rm gas}} = 0.750$ for the baryon.
The mean angle between the two angular momentum vectors 
is $\theta_{\rm ave} \sim 28.2^\circ$ and the median is $\theta_{\rm med} \sim 19.1^\circ$.

\begin{figure}
\begin{center}
\resizebox{6cm}{!}{\includegraphics[clip]{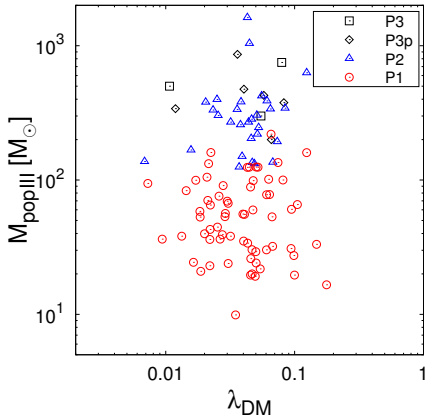}}
\caption{
The final stellar masses are plotted versus spin parameters of the parent dark matter halos 
to demonstrate the degree of correlation between the two quantities.
}
\label{fig:Spin-Mstar}
\end{center}
\end{figure}

Recently, \cite{desouza13} developed an analytic model assuming a tight correlation 
between the stellar mass and the spin parameter of the host halo.
They argue that the mass distribution of the first stars would be shaped by 
the angular momentum distribution of the host halos 
that is well described by a log-normal function.
We have discussed in section 4 that 
rotational support delays or possibly prevents cloud collapse
and effectively reduces the accretion rate onto the central protostar. 
Our results, however, show a weaker correlation between 
the spin parameter and the final stellar mass (see Fig. \ref{fig:Spin-Mstar})
than the dependence on the other two parameters, $z_{\rm form}$ and $M_{\rm virial}$.
\cite{desouza13} use the semi-analytic model of \cite{mckee08} 
to calculate the stellar mass starting from the angular momentum of the host halos. 
It is important to point out that \cite{mckee08} use 
$f_{\rm Kepler}$ of a gas cloud, rather than that of a halo, in their
accretion model. Although the parent gas cloud properties almost directly
affect the pre-stellar collapse, it is less clear how the spin of
dark halos affect the {\it small-scale} processes. Indeed we find
no clear trend and thus the stellar mass cannot be estimated from the host halo spin. 

\subsection{Varieties of Sub-structure of Star-forming Clouds}

The variety of mass accretion histories and hence the final stellar masses 
likely originate from different initial conditions during formation of gas clouds. 
We have already shown that the overall shapes and structures of collapsing gas clouds 
reflect different initial conditions. 
With more than one hundred sample members in our simulations, 
we find a few cases which show very different properties. 
In the following we discuss these clouds in greater detail.

\subsubsection{The Most Rapidly and the Most Slowly Rotating Clouds}

\begin{figure}
\begin{center}
\resizebox{7.5cm}{!}{\includegraphics[clip]{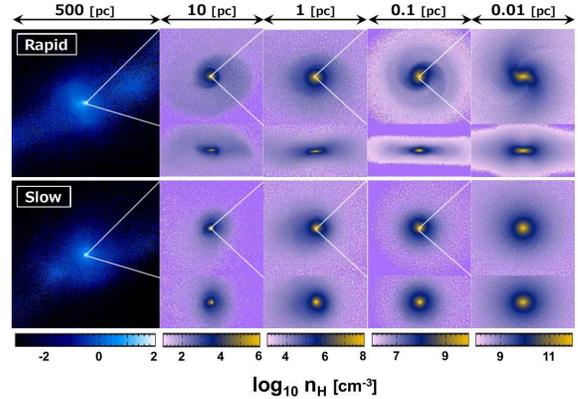}}
\caption{
Projected gas density distributions 
when the central density reaches $10^{12} \ {\rm cm^{-3}}$ for two extreme cases: 
the most rapidly ($top$) and the most slowly ($bottom$) rotating clouds. 
The physical properties of theses clouds are summarized in Table \ref{table:abnormal}.
}
\label{morphology:extremely}
\end{center}
\end{figure}

Rotation of a cloud plays 
a critical role in determining the thermal evolution during the pre-stellar collapse. 
Figure \ref{morphology:extremely} presents the density distributions 
at multiple epochs for the two extreme cases of very rapidly and very slowly rotating clouds 
($top$ and $bottom$ panels; Table \ref{table:abnormal} shows their properties). 
With fast rotation,
the cloud has a disk-like structure that is nearly rotationally supported. 
Interestingly, this particular cloud is prominent also in the other figures; 
it shows the lowest temperature evolution track in Figure \ref{plot_PROFILE_temp_multi} and 
the slowest accretion history in Figure \ref{plot_EXTEL_multi}. 
The cloud collapses slowly because of rotational support, 
and cools down to $\simeq 50 \ {\rm K}$, 
almost reaching the CMB temperature floor at 
$T_{\rm CMB} \sim 2.73 (1+z) \ {\rm K}$. 
The resulting final stellar mass is only $9.9 \ {\rm M_{\odot}}$, 
which is the smallest among our samples. 

The opposite limiting case of the most slowly rotating cloud 
is presented in the lower panels of Figure \ref{morphology:extremely}. 
The cloud shows little indication of a flattened structure and
almost no difference between the face-on and edge-on views. 
This case corresponds to the case with the lowest value of $f_{\rm Kepler}$ 
plotted in Figure \ref{plot_PROFILE_Menc-Fkep}.
It has $f_{\rm Kepler} \simeq 0.2$, less than half of the average.
Without strong rotational support, the cloud quickly collapses and the
gas mass accretion rate onto the central protostar is large,
yielding the final mass of $\simeq 100 \ {\rm M_{\odot}}$ 
(The slowest rotating cloud does not form the most massive star 
because the protostellar evolution depends also on the
cloud mass at the onset of collapse, $M_{\rm cloud}$ ; 
see Figure \ref{plot_LIST_Estimation-MpopIII}a in Section 4.2).
Clearly, nearly spherical accretion is a favored condition
for the formation of very massive primordial stars.

\begin{deluxetable}{crccrl}
\tablewidth{0pt}
\tablenum{4}
\tablecaption{Cloud's Properties of Two Exceptional Cases}
\tablehead{
\colhead{Case} & 
\colhead{$M_{\rm popIII}$} &
\colhead{$z_{\rm form}$} & 
\colhead{$M_{\rm virial}$} &
\colhead{$M_{\rm cloud}$} &
\colhead{$\beta_{\rm cloud}$} \\
\colhead{} & 
\colhead{$({\rm M_{\odot}})$} &
\colhead{} & 
\colhead{$(10^5 \ M_{\odot})$} &
\colhead{$({\rm M_{\odot}})$} &
\colhead{}}
\startdata
Rapid &  9.9 & 14.90 & 5.24 & 25 & 0.579 \\
Slow & 105.0 & 13.80 & 2.86 & 278 & 0.088
\enddata
\tablecomments{Each column header has the same meaning as in Table \ref{table:average}.}
%\tablerefs{}
\label{table:abnormal}
\end{deluxetable}

\subsubsection{Multiple Density Peaks}

\begin{figure}
\begin{center}
\resizebox{7.5cm}{!}{\includegraphics[clip]{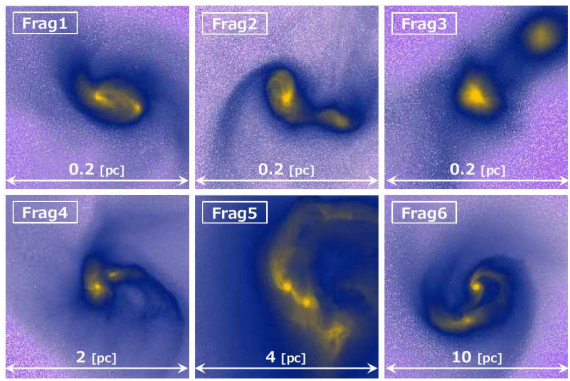}}
\caption{
Projected gas density distributions for six clouds which have multiple density peaks 
when the central density reaches $10^{12} \ {\rm cm^{-3}}$. 
The clump properties are summarized in Table \ref{table:fragment}.
}
\label{morphology:fragment}
\end{center}
\end{figure}

Some of our samples show the formation of multiple density peaks
in and around a single gas cloud.
We have already seen their signatures in Figure \ref{plot_PROFILE_Radi-Dens}
as humps in the radial density profiles. 
We find seven close density peaks in six clouds. 
Figure \ref{morphology:fragment} shows the projected gas density distributions around them.
Note that ``Frag5'' has two neighboring clumps. 
Several properties of the neighboring gas clumps are summarized in Table \ref{table:fragment}. 
The clumps are located at distances $0.05 \sim 2 \ {\rm pc}$ from the central collapsing core. 
Note that we only follow the protostellar evolution of the central core. 
The clumps in the surrounding are smeared out in the subsequent 2D radiation hydrodynamic 
simulations by averaging the physical quantities over the azimuthal direction. 
Therefore, whether or not stars can be formed in the neighboring clouds remains 
unclear in our calculations. Fully three dimensional calculations
of protostellar evolution are necessary to determine the final stellar mass(es) for these cases.

We mention that the cosmological simulation of \cite{turk09} also show two fragments 
forming with a $800 \ {\rm AU}$ separation during the run-away collapse of a cloud. 
We find similar multiple clumps but with even wider separations, 
though only in a handful of cases out of our sample of 110 clouds. 
Note also that the clump formation is different from those seen in a self-gravitating circumstellar disk 
\citep[e.g.,][and also see the next section]{stacy13b}, 
which produce small separation ($< 1000$ AU) multiple protostellar systems. 
Recently, \cite{greif13} show the possibility of the chemothermal instability which makes fragments on a scale of a few tens of au.

\begin{deluxetable}{ccrlrr}
\tablewidth{0pt}
\tablenum{5}
\tablecaption{Properties of Seven Gas Clumps}
\tablehead{
\colhead{No.} & 
\colhead{$R$} &
\colhead{$M(<R/2)$} &
\colhead{$\rho_{\rm cen,frag}$} &
\colhead{$t_{\rm ff,frag}$} &
\colhead{$M_{\rm popIII,prim}$} \\
\colhead{} & 
\colhead{$({\rm pc})$} & 
\colhead{$({\rm M_{\odot}})$} &
\colhead{$({\rm cm^{-3}})$} &
\colhead{$({\rm year})$} &
\colhead{$({\rm M_{\odot}})$}}
\startdata
1 & 0.05 & 170 & $1.48 \times 10^{10}$ & 436 & 380.3 \\
2 & 0.08 & 50 & $1.92 \times 10^8$ & 3820 & 53.2 \\
3 & 0.10 & 170 & $2.58 \times 10^8$ & 3300 & 381.3 \\
4 & 0.20 & 120 & $2.30 \times 10^7$ & 11100 & 340.0 \\
5a & 0.60 & 270 & $6.59 \times 10^7$ & 6530 & 27.4 \\
5b & 1.50 & 900 & $3.57 \times 10^5$ & 88700 & 27.4 \\
6 & 2.00 & 1300 & $7.20 \times 10^5$ & 62500 & 55.5 \\
\enddata
\tablecomments{
Column 2: Distance between the primary star and the clump, 
Column 3: Enclosed mass within $R/2$, 
Column 4: Central density of the clump, 
Column 5: Estimated free-fall time, and
Column 6: The final stellar mass of the primary star. 
}
%\tablerefs{}
\label{table:fragment}
\end{deluxetable}

\subsection{Accretion Disk Fragmentation}

Recently, \cite{clark11} and \cite{greif11} studied the
evolution of primordial protostellar disks over one hundred years using sink particle techniques. 
In their simulations, multiple protostars are formed in a disk via gravitational instability.
Although our two-dimensional calculations cannot follow the evolution
of multiple protostars in a disk, it is worth discussing the possible effect and outcome in such cases.

If disk fragmentation occurs during the mass accretion phase, 
there are two mechanisms that can reduce the final stellar mass of the central main protostar.
One is the reduction of the total gas mass that can be accreted onto the central
star, simply because additional sinks are present
\citep[fragmentation-induced starvation; i.e.,][]{peters10}.
The other is that the structure and evolution of the protostar itself 
are affected by the reduced accretion rate in a complicated nonlinear
way, as we have discussed in the previous sections.
With a lower accretion rate, the protostar begins KH-contraction at a lower 
stellar mass so that the mass accretion is halted by UV radiative 
feedback in accordance to our P1 evolution. 
We expect that both of these effects will reduce the final mass of the central protostar.

There is another physical process that we have not included.
Primordial magnetic fields are generally considered to be much weaker 
than in the present-day star-forming cloud, and thus
their effects are often ignored in the study of primordial star formation.
However, as shown by \cite{machida10},  
even $pico$-$gauss$ magnetic fields are sufficient to 
transfer angular momentum in the accretion flow via magnetic braking.
The accretion disk disappears in cases with a weak magnetic field, and  
the surrounding gas is accreted onto the protostar directly in a roughly spherical manner. 
More recently, \cite{turk12} and \cite{sur12} perform magneto-hydrodynamic simulations
of primordial star formation. They show that 
turbulent velocity fluctuations resolved in their simulations
can amplify the magnetic field via the dynamo effect.
They argue that magnetic fields can influence the formation 
and the structure of the accretion disk. Further studies on the effects
of magnetic fields are needed to determine the overall
impact on the characteristic masses of the first stars.

\begin{figure}
\begin{center}
\resizebox{7cm}{!}{\includegraphics[clip]{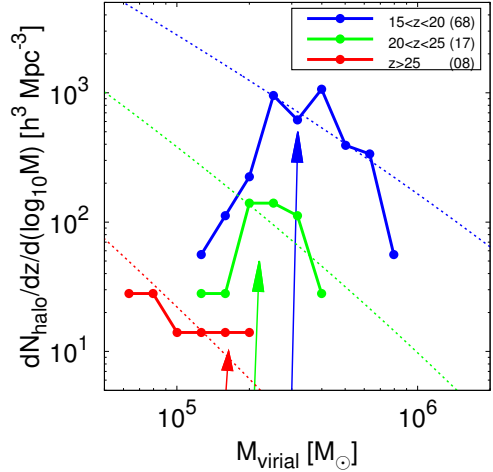}}
\caption{
Mass functions of the dark halos from our simulations (solid lines). 
The dotted lines represent the analytic mass functions 
calculated with the public code developed by \cite{reed07}. 
Different colors represent the different formation redshifts. 
Note that the solid lines have been shifted vertically for comparison.
The arrows show the characteristic halo masses collapsing 
at redshift 17.5 (blue), 22.5 (green), and 27.5 (red), respectively 
\citep[using Eq. 26 in][for $T_{\rm vir} \sim 800 \ {\rm K}$]{barkana01}.
}
\label{plot_IMFpred_Mvir-dNdz}
\end{center}
\end{figure}

\subsection{Distribution of Stellar Masses}

Our study shows a wide mass range 
for the first generation of stars (Figure \ref{plot_IMF-PopIII_3PATHp}),
including several very massive stars with
$M_{\rm popIII} > 300 \ {\rm M_{\odot}}$.
However, the sample should not be regarded as a complete set,
because our halo selection is somewhat arbitrary, and
the total simulation volume is still too small to be
representative of a {\it cosmological} volume. 
In this section, we first examine if and how our halo selection
is biased and then discuss the fate of the first stars.

\subsubsection{Sampling Bias}

We use a zoom-in technique to achieve extremely high resolution 
in large simulation-boxes of $L_{\rm box} = 1$ and $2 \ {\rm h^{-1} \ Mpc}$.
Although we choose many halos,
other (unselected) halos are present that can possibly host first stars.
Also, the largest volume of our simulations is $(2 \ {\rm h^{-1} \ Mpc})^3$.
We still miss very rare objects which should exist in a much larger (by a factor $\sim 10^8$)
{\it cosmological} volume of $(1 \ {\rm h^{-1} \ Gpc})^3$. 
The earliest forming halo in our sample is located at $z \sim 35$,
whereas the very first star in the observable volume is expected to form 
at $z \to 50$ \citep{gao07, naoz05}. 

We calculate the mass function of our sample dark matter halos that host primordial stars, 
and compare it to the Press-Schechter mass function. 
Figure \ref{plot_IMFpred_Mvir-dNdz} compares 
the halo mass functions $dN_{\rm halo}/dz/d(\log_{10}M)$ at different redshifts. 
Note that we are concerned with only the overall shape of the mass functions 
because, with our small, $\sim$ Mpc volume, 
the absolute abundance of first star hosts cannot be evaluated robustly.
Thus, the amplitudes are shifted arbitrarily for comparison.
Three dotted lines in Figure \ref{plot_IMFpred_Mvir-dNdz} are
the Press-Schechter mass functions calculated by the public code of \cite{reed07}.
The lower cut-off of the solid lines are a physical effect,
reflecting the critical collapse mass at the respective redshift. 
Note that this is not the resolution limit; 
the DM particle mass is $< 20 \ {\rm M_{\odot}}$ and
thus the effective mass resolution is much lower than the cut-off at $\sim 10^4 \ {\rm M_{\odot}}$, 
which is resolved by more than 5000 N-body particles. 
Overall, the mass function of our sample halos represents 
the high-mass end and is roughly consistent with the underlying true mass function.
Considering the dependence of the final stellar mass on collapse redshift and halo mass, 
we conclude that the derived stellar mass distribution is not significantly biased.

\subsubsection{Final Fate of the First Stars}

The main result of our calculated mass distribution indicates that the first stars
will end their lives in a variety of ways.
Table \ref{t3} lists the final fate of 
zero-metallicity stars calculated by \cite{yoon12}. 
There are two sets for different rotational degrees 
$f_{\rm Kepler} = 0$ (non-rotating) and $0.5$ (rapid-rotating).
The corresponding fate for different initial stellar mass ranges are given there.
In the table, $N_{\rm sample}$ is the number of stars in the respective
mass range found in our simulations.
A large fraction of them (about 60\%) leave massive black-holes (BH),
whereas 11\% become neutron stars (remnant of CCSN) and
20\% die as pair-instability supernovae regardless of the rotational degrees.
We note that there are basically two channels for the 
Pop III stars to disperse the first heavy elements synthesized in them;
core-collapse supernovae and pair-instability supernovae,
which leave different characteristic elemental yields.

\begin{table}[t]
\caption{Final Fate of Pop III Stars\label{t3}}
\begin{center}
\begin{tabular}{cr@{$M$}lcr@{$M$}lc}
\tableline\tableline\noalign{\smallskip}
Case & \multicolumn{2}{c}{$M_{\rm ZAMS} \ ({\rm M_{\odot}})^1$} & $N_{\rm sample}$ & \multicolumn{2}{c}{$M_{\rm ZAMS} \ ({\rm M_{\odot}})^1$} & $N_{\rm sample}$ \\
 & \multicolumn{3}{c}{(non-rotating)} & \multicolumn{3}{c}{($f_{\rm Kepler, ZAMS} = 0.5$)} \\
\noalign{\smallskip}\tableline\noalign{\smallskip}
NS    &   8 $<$ & $<$  25 & 12 &   8 $<$ & $<$  25 & 12  \\
BH    &  25 $<$ & $<$  80 & 36 &  25 $<$ & $<$  65 & 28  \\
PPISN &  80 $<$ & $<$ 120 & 10 &  65 $<$ & $<$  90 & 10  \\
PISN  & 120 $<$ & $<$ 240 & 21 &  90 $<$ & $<$ 200 & 24  \\
BH    & 240 $<$ &         & 31 & 200 $<$ &         & 36  \\
\noalign{\smallskip}\tableline\noalign{\smallskip}
\end{tabular}
\tablecomments{
Column 1: final fate of stellar evolution (neutron star (NS), black hole (BH), pulsation pair-instability supernovae (PPISN), and pair-instability supernovae (PISN)), 
Column 2 and 4: stellar mass range at ZAMS, and
Column 3 and 5: number in our sample.
Column 2 and 3 are for the non-rotating star, whereas 
column 4 and 5 are for the highly rotating star.
}
\tablerefs{(1) From Figure 12 in Yoon et al. 2012.}
\end{center}
\end{table}
%%%
%\begin{deluxetable}{cr@{$M$}lcr@{$M$}lc}
%\tablewidth{0pt}
%\tablenum{6}
%\tablecaption{Final Fate of Pop III Stars}
%\tablehead{
%\colhead{Case} &  
%\multicolumn{2}{c}{\colhead{$M_{\rm ZAMS} \ ({\rm M_{\odot}})^1$}} & 
%\colhead{$N_{\rm sample}$} &
%\multicolumn{2}{c}{\colhead{$M_{\rm ZAMS} \ ({\rm M_{\odot}})^1$}} & 
%\colhead{$N_{\rm sample}$} \\
%\colhead{} &  
%\multicolumn{3}{c}{\colhead{(non-rotating)}} &
%\multicolumn{3}{c}{\colhead{($f_{\rm Kepler, ZAMS} = 0.5$)}}} 
%\startdata
%NS    &   8 $<$ & $<$  25 & 12 &   8 $<$ & $<$  25 & 12  \\
%BH    &  25 $<$ & $<$  80 & 36 &  25 $<$ & $<$  65 & 28  \\
%PPISN &  80 $<$ & $<$ 120 & 10 &  65 $<$ & $<$  90 & 10  \\
%PISN  & 120 $<$ & $<$ 240 & 21 &  90 $<$ & $<$ 200 & 24  \\
%BH    & 240 $<$ &         & 31 & 200 $<$ &         & 36  \\
%\enddata
%\tablecomments{
%Column 1: final fate of stellar evolution (neutron star (NS), black hole (BH), pulsation pair-instability supernovae (PPISN), and pair-instability supernovae (PISN)), 
%Column 2 and 4: stellar mass range at ZAMS, and
%Column 3 and 5: number in our sample.
%Column 2 and 3 are for the non-rotating star, whereas 
%column 4 and 5 are for the highly rotating star.
%}
%\tablerefs{(1) From Figure 12 in Yoon et al. 2012.}
%\label{t3}
%\end{deluxetable}
%%%

No clear signature of PISNe is found so far
observationally. Although we see a slight deficit in the
PISN mass range (Fig. 5), the obtained mass distribution is sufficiently broad
that a significant number of PISN very likely occurred in the early universe. 
Thus, it is necessary to explain the apparent
paucity of the PISN signatures in, for example, the elemental
abundance patterns of Galactic metal-poor stars.
First, there could be some caveat in our calculations, such as
our two-dimensional treatment of accretion, which precludes fragmentation of the accretion disks. 
Our selection of halos to investigate may miss a large number of
lower mass halos and/or so-called Pop III.2 stars (see below).
Second, there may be possible biases in observations.
The current survey of metal-poor stars are targeted halo stars, 
whereas most of the remnants of the first stars could be concentrated 
toward the central regions of the Galaxy \citep{white00, tumlinson10}.
PISNe can enrich the surrounding gas promptly to a large
metallicity, and thus surveys of metal-poor stars
might actually miss the stars with PISN signatures \citep{karlsson08, karlsson13}. 

In the present paper, we have only considered the so-called Pop III.1 stars, 
for which the initial conditions are completely determined cosmologically. 
There are other populations of primordial stars, called Pop III.2 stars, 
which are formed in a primordial gas 
that has been affected by the prior existence of other stars
%%%[SH131125 : reply for the referee's 2nd comment]
\citep[both the Pop III and Pop II stars; the latter can contribute orders of magnitude more strongly, in particular, at low redshifts, $z < 15$,  as shown in Figure 6 of][]{agarwal12}
%%%
. 
These additional populations are characterized by 
the differences of physical conditions where they are formed, 
for example inside H{\sc ii} regions (Pop III.$2_{\rm ION}$) and
inside photo-dissociation regions (Pop III.$2_{\rm DIS}$).
Pop III.2 stars are thought to have lower masses 
on the average than Pop III.1 stars 
\citep{yoshida07, hosokawa12b}.
Determining the primordial stellar IMF must necessarily involve 
all of these cases in a fully cosmological context. 
It is however beyond the scope of the present paper.

\section{Conclusion}

To study the mass distribution of the first stars, 
we have performed more than hundred simulations of first star formation 
in a proper cosmological context.
The resulting stellar masses range from 
$M_{\rm popIII} \sim$ 10 to 1000 ${\rm M_{\odot}}$. 
Most of them are distributed around a few tens to a few hundreds solar masses. 
A number of the first stars would end their lives leaving massive BHs, 
and a significant fraction of stars die as
supernovae. We summarize our main results as follows.

\begin{enumerate}
\item 
The collapse time-scale is the key quantity that determines
the thermal evolution of the gas clouds.
Rapidly collapsing clouds follow the well-known evolution 
path which is controlled by ${\rm H_2}$ cooling. 
Slowly collapsing clouds cool efficiently 
because of the less effective compressional heating and the
longer time scale for the coolants to be produced.
When the temperature decreases to $\sim 100$ Kelvin, HD molecules are formed efficiently, and 
the additional HD molecular cooling further lowers the temperature to the CMB temperature. 
The physical structure of the gas envelope around 
a protostellar core also differs among our sample of clouds, 
reflecting the different thermal evolution during the collapse. 
This in turn leads to different accretion histories 
onto the protostars as seen in the subsequent evolutionary stage. 
\item 
The main self-regulation mechanism of protostellar growth
is UV feedback; the resulting stellar mass differs considerably
among our sample.
On average, the final stellar mass is larger for more 
rapid mass accretion. Protostars approach the ZAMS
at larger stellar masses in such cases.
Interestingly, in cases with very rapid accretion 
$\dot{M} \gtrsim 4 \times 10^{-3} \ {\rm M_{\odot} \ yr^{-1}}$, 
the protostar greatly expands so that the stellar 
effective temperature remains below $10^4$ K. 
The ionizing photon luminosity remains so low 
that UV feedback never becomes strong enough to prevent
mass accretion \citep[e.g.,][]{hosokawa12a}. 
We argue that rapid mass accretion is a promising path
for forming very massive ($\gtrsim 10^3 \ {\rm M_{\odot}}$) 
stars in the early universe \citep[see e.g.][]{hosokawa12a, hosokawa13}.
\item 
The overall accretion rate can be estimated from the structure 
of the gas envelope around a protostellar core.
The structure itself is affected by the 
thermal evolution during the run-away collapse.
We can predict the final stellar mass from the conditions
in the early state of the gas cloud formation,
and to a less rigorous extent even from the properties of dark matter halos.
In fact, we have seen correlations between the final stellar 
masses and infall rates at the scales of both the gas clouds 
and the dark matter halos (Eqs. \ref{eq:MpopIII-dMdtJeans} and \ref{eq:MpopIII-dMdtVirial}). 
We have identified key physical quantities that determine the infall rates: 
the parent cloud mass and the rotation parameter (for the cloud scale), 
and the formation redshift and the virial mass of the host halo (for the halo scale).
\end{enumerate}

Through these simulations we have found new protostellar 
evolutionary paths (P2 and P3) that typically lead to
the formation of very massive ($>$ a few 100 ${\rm M_{\odot}}$) first stars.
Although this mode of first star formation is not dominant,
the relative abundance of the most massive stars is important
for the study of the formation of supermassive black holes in the early universe.

We have identified a key feedback mechanism that
regulates the mass growth and determines the final 
mass of a primordial protostar.
Because the strength of the radiative feedback is 
determined by the protostellar evolution itself,
accurate modeling of the time-dependent accretion  
and the resulting protostellar evolution is clearly needed.
Statistical studies using 3D radiation hydrodynamic simulations 
would be one of the next challenges for deriving 
the initial stellar mass function of the first stars.

We have shown that 
the final mass of a central star can be estimated reasonably well from  
a few physical properties of the host halo.
In principle, we can derive the ``cosmological'' stellar IMF theoretically 
if we know the mass function of the host dark matter halos and its redshift-evolution, 
and also the other important quantities such as the gas infall rates.
The results can be used, for example, to predict the early chemical 
evolution and the characteristic signatures of the first galaxies. 
It is definitely worth developing such a global evolution model
using large-scale cosmological simulations.

Future observations will exploit the next-generation facilities
such as JWST, 30 - 40 meter telescopes, and radio telescope arrays,
to probe the evolution of the high-redshift universe.
Data from such observations, combined with detailed studies of the 
Galactic metal-poor stars, will ultimately constrain the characteristic
mass of the first generation of stars.

\acknowledgments 

We are grateful to Sanemichi Z. Takahashi,  for discussions on the numerical treatment of angular momentum transport.
We also thank Masahiro N. Machida, Hajime Susa, Kenji Hasegawa, Kei Tanaka, and Kohei Inayoshi for stimulating discussions 
and Thomas H. Greif for helpful comments on this study.
The numerical calculations were in part carried out on 
T2K-Tsukuba System at Center for Computational Sciences, University of Tsukuba, 
SR16000 at YITP in Kyoto University, and
Cray XT4, XC30 and the general-purpose PC farm at Center for Computational Astrophysics, CfCA, of National Astronomical Observatory of Japan.
The work is supported in part by 
the Global COE Program ``The Physical Sciences Fron-tier,'' MEXT, Japan (SH), 
the Grands-in-Aid by the Ministry of Education, Science and Culture of Japan (25800102 TH, 21684007 and 25287040 KO), and 
a grant from the Hayakawa Satio Fund awarded by the Astronomical Society of Japan. 
Portions of this work were conducted at the Jet Propulsion Laboratory, 
California Institute of Technology, operating under a contract with the 
National Aeronautics and Space Administration (NASA). 
The page charge of this paper is partly supported by Center for Computational Astrophysics, National Astronomical Observatory of Japan.

%=== References ===%

\bibliography{biblio}
\bibliographystyle{apj}

%=== Appendix ===%
\appendix

\section{Angular Momentum Transport in the Accretion Disk}

In a rapidly accreting circumstellar disk, angular momentum is transfered 
outward by torques produced by non-axisymmetric spiral structure.
In our previous work \citep{hosokawa11}, 
we adopted the so-called $\alpha$-viscosity \citep{shakura73} 
to mimic this effect in axisymmetric 2D radiation hydrodynamic simulations. 
The equation of angular momentum transport is written
\begin{equation}
\frac{\partial A}{\partial t} + \nabla \cdot (A \vec{v}) = - \frac{1}{R} \frac{\partial}{\partial R} \left( R^3 \eta \frac{\partial \Omega}{\partial R} \right) \ ,
\label{eq:ang-transport}
\end{equation}
where $\Omega$ is angular velocity, 
$\eta = 2 \alpha \rho c_s^2 / (3 \Omega)$, and 
$\alpha$ is a dimensionless free parameter.
\cite{hosokawa11} model the spatial distribution of the alpha-parameter with an analytic function
\begin{equation}
\alpha(R, Z) = \alpha_0 \exp \left( - \frac{Z}{H(R)} \right) \ ,
\label{eq:alpha-height}
\end{equation}
where $\alpha_0$ is a constant free parameter, and 
$H(R)$ is the scale height of the circumstellar disk at each radial coordinate $R$. 
The fiducial value of $\alpha_0$ is 0.6 in \cite{hosokawa11}.

In this paper, we improve the above description with an additional $R$-dependence of $\alpha_0$. 
Following \cite{zhu10b} and \cite{takahashi13}, 
we adopt the functional form proposed by \cite{gammie96}, 
\begin{equation}
\alpha_0(R) = \alpha_{\rm max} \ e^{- Q(R)^4} \ ,
\label{eq:alpha-q}
\end{equation}
where $Q(R)$ is the Toomre $Q$-parameter \citep{toomre64}, 
which measures the gravitational stability of the disk. 
The underlying idea of the above dependence is as follows. 
With the small values of $Q < 1$, 
$\alpha_0$ approaches $\alpha_{\rm max}$ 
because the angular momentum transport should be efficient 
with spiral arms emerging in such a highly unstable disk. 
The large $\alpha_0$ promotes the mass accretion, 
which reduces the disk surface density and increases $Q$. 
On the other hand, $\alpha_0$ has a cut-off for $Q > 1$, 
because the spiral arms disappear in a stable disk. 
Materials falling onto the disk accumulate without angular momentum transport, 
which enhances the disk surface density and reduces $Q$. 
With the above regulation mechanism, 
the self-gravitating disk hovers around the marginally stable steady state with $Q \sim 1$.  

\begin{figure}
\begin{center}
\resizebox{6cm}{!}{\includegraphics[clip]{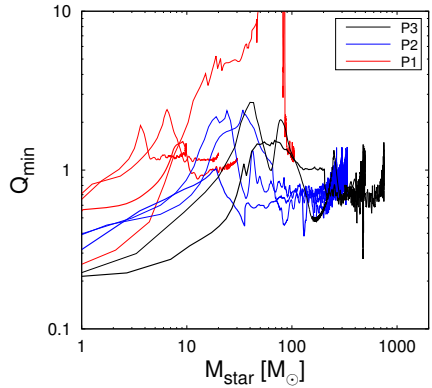}}
\caption{
Evolution of the minimum Toomre $Q$-parameter as a function of the stellar mass 
for the same cases as in Figure \ref{plot_PROFILE_Radi-Dens}. 
We see that $Q_{\rm min}$ is close to unity, which is generally expected for 
a self-gravitating circumstellar disk, in almost all the cases. 
Only one case, for which the cloud rotation is the slowest of the entire sample, 
shows a very large $Q_{\rm min}$ (see discussion in text).
}
\label{plot_EXTEL_Mstr-Qval}
\end{center}
\end{figure}

In this paper, we adopt $\alpha_{\rm max} = 2$ for all the examined cases. 
Figure \ref{plot_EXTEL_Mstr-Qval} shows the evolution of the minimum Toomre $Q$-values in several cases. 
We see that $Q_{\rm min}$ eventually converges to around unity in each case. 
Only one case shows a peculiar evolution with $Q_{\rm min}$ reaching 10 at maximum. 
This corresponds to the most slowly rotating cloud explained in Sec. 5.2.1. 
A circumstellar disk hardly forms in this case. 

Mass accretion continues smoothly 
without ``ring-like'' disk fragmentation of the disk in our simulations. 
In test calculations with lower values of $\alpha_{\rm max}$
ring-like fragmentation does occur occasionally with very rapid mass accretion. 
We avoid this on purpose with our choice of $\alpha_{\rm max}$
because we cannot treat fragmentation in 2D axisymmetric simulations. 
We consider our results for very rapid mass accretion 
a conservative upper limit of the final stellar mass resulting from stellar UV feedback. 
Disk fragmentation could further reduce the final stellar masses 
(also see Sec. 5.3).

\section{Modeling Protostellar Evolution}

As described in Sec. 2.2, 
our stellar evolution code occasionally has convergence difficulties
for the extremely high and highly variable accretion rates encountered
in the P2 and P3 scenarios.  Whereas for the P1 cases
we can use the same method as \cite{hosokawa11}, we switch to 
the following simple procedure of protostellar evolution when convergence
difficulties arise.

\subsection{Oscillating Protostar (P2)}

\begin{figure}
\begin{center}
\resizebox{7cm}{!}{\includegraphics[clip]{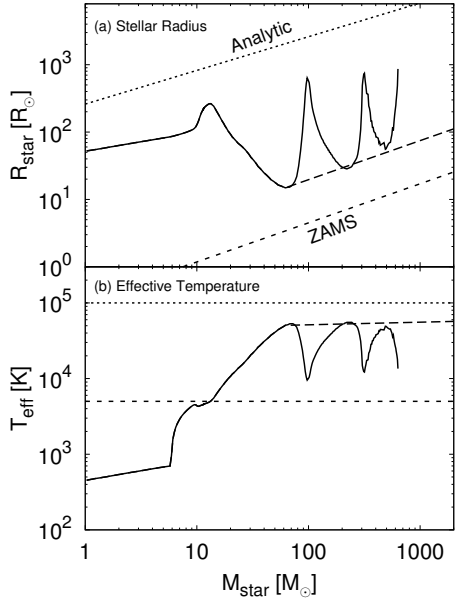}}
\caption{
Our analytic model for the evolutionary path P2 of 
an accreting protostar (long-dashed lines). 
The top and bottom panels show the evolution of 
the stellar radius and effective temperature with increasing the stellar mass. 
The solid lines show the numerical results for 
$\dot{M} = 6 \times 10^{-3} \ {\rm M_{\odot}} \ {\rm yr^{-1}}$ 
taken from \cite{hosokawa12a}. 
The dotted and short-dashed lines are the same as in 
Figures \ref{plot_EXTEL_Mstr-dMdt-wowi}(b) and (c).
}
\label{plot_EXTEL_history_P2}
\end{center}
\end{figure}

The solid line in Figure \ref{plot_EXTEL_history_P2} shows 
the numerically calculated evolution with the constant accretion rate 
$\dot{M} = 6 \times 10^{-3} \ {\rm M_{\odot}} \ {\rm yr^{-1}}$. 
In this case, our stellar evolution code experienced convergence difficulties 
after the stellar mass exceeded $60 \ {\rm M_{\odot}}$, 
when the radius began to increase and the
total luminosity reached the Eddington limit
(also see Sec. 2.2.1). 
The problem worsened when we used the time-dependent mass accretion histories obtained in the simulations. 

We adopt the following simplified procedure instead. 
First, we numerically calculate the protostellar evolution 
until the protostar begins to oscillate. 
After that, 
we analytically model the evolution of the stellar radius and luminosity 
using the same power-law functions as for non-accreting ZAMS stars,
\begin{eqnarray}
R(M_*) =& R_{\rm edd} \left( \frac{M_*}{M_{\rm edd}} \right)^{0.58} \ {\rm R_{\odot}} \ , \\
L(M_*) =& L_{\rm edd} \left( \frac{M_*}{M_{\rm edd}} \right)^{1.30} \ {\rm L_{\odot}} \ ,
\label{eq:P2eq1}
\end{eqnarray}
where $M_{\rm edd}$, $R_{\rm edd}$, and $L_{\rm edd}$ are 
the stellar quantities when the star begins to oscillate. 
Once the accretion rate drops below the critical value 
$4 \times 10^{-3} \ {\rm M_{\odot}} \ {\rm yr^{-1}}$ (Eq. \ref{eq:P2-criterion}), 
we let the star to contract over the KH timescale according to
\begin{eqnarray}
R(M_*) =& R_{\rm prev} + dt \cdot {(R(M_*)_{\rm ZAMS} - R_{\rm prev})/t_{\rm KH}} \ {\rm R_{\odot}} \ , \\
L(M_*) =& L_{\rm prev} + dt \cdot {(L(M_*)_{\rm ZAMS} - L_{\rm prev})/t_{\rm KH}} \ {\rm L_{\odot}} \ ,
\label{eq:P2eq2}
\end{eqnarray}
where $R_{\rm prev}$ and $L_{\rm prev}$ are 
the stellar quantities at the previous step of calculation, and
\begin{eqnarray}
R(M_*)_{\rm ZAMS} =& 3.109 \times 10^{-1} \ \left( \frac{M_*}{M_{\rm edd}} \right)^{0.58} \ {\rm R_{\odot}} \ , \\
L(M_*)_{\rm ZAMS} =& 3.939 \times 10^3 \ \left( \frac{M_*}{M_{\rm edd}} \right)^{1.30} \ {\rm L_{\odot}} \ .
\label{eq:P2eq3}
\end{eqnarray}
After the star reaches the ZAMS stage, 
the radius and luminosity are assumed to be the same as those of non-accreting ZAMS stars.

\subsection{Super-Giant Protostar (P3)}

The convergence difficulties described above worsen when considering P3 evolution.
In this case we first perform the 2D RHD simulation without stellar UV feedback and 
record the mass accretion history without calculating stellar evolution.
Next, we numerically calculate stellar evolution as a post process from the accretion history. 
%We here employ a different stellar evolution code \citep{hirano11} 
%for avoiding the convergence difficulties. 
We record the stellar evolution track until the stellar total luminosity reaches the Eddington value. 
We repeat the RHD simulation and include UV feedback using the recorded stellar evolution track. 
Finally, we switch to the same analytic model as for the path P2 
for the evolution beyond the recorded stellar evolution track. 
We find that UV feedback significantly affects the accretion
only after the protostar enters the evolutionary path P2.

It would be very time-consuming to use the procedure 
described above for all twelve cases of P3 evolution. 
Instead, we have used this procedure for only three cases. 
For the other nine cases, we estimate the final stellar masses using the relation (\ref{eq:Mest-Jeans}). 
We designate these nine cases as the path P3p in the captions of Figures \ref{plot_IMF-PopIII_3PATHp} and \ref{plot_EXTEL_multi}.

\section{Dependence of thermal evolution on the collapsing timescale}

\begin{figure}
\begin{center}
\resizebox{7.5cm}{!}{\includegraphics[clip]{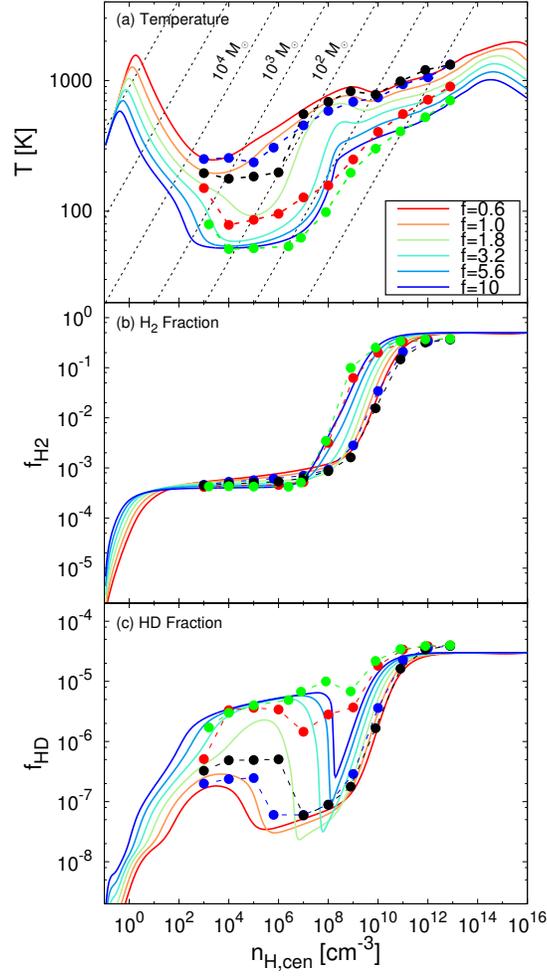}}
\caption{
Comparison of the thermal evolution during the cloud collapse 
between the simulation results (dashed lines and filled circles) and one-zone models (solid lines). 
The solid lines with the different colors represent the different collapse timescales, 
$t_{\rm coll} = f \cdot t_{\rm ff}$, with $f$ = 0.6, 1.0, 1.8, 3.2, 5.6, 10, 
which is a parameter for one-zone modeling. 
The red, blue, and black dashed lines represent the same 3D simulation cases shown in Figure \ref{morphology:classification}, 
whereas the green line is for the most rapidly rotating cloud shown in Figure \ref{morphology:extremely}. 
The black dashed lines in the top panel present $\rho$ - $T$ relations for given values of the Jeans mass.
}
\label{plot_100FS_one-zone3}
\end{center}
\end{figure}

As described in Sec. 3.2.2, HD line cooling, 
which has been thought to be unimportant in Pop III.1 star formation, 
is actually quite effective during the early run-away collapse stage 
in several of our cases. 
We argue that HD molecules form 
when rotational support of the cloud can sufficiently slow the collapse. 
Here we test this hypothesis using a simple one-zone model, 
which follows the thermal evolution at the center of the collapsing cloud 
\citep[e.g.,][]{omukai00, chiaki13}. 
Figure \ref{plot_100FS_one-zone3} shows the results for different collapsing timescales, 
$t_{\rm coll} = f \cdot t_{\rm ff}$, with $f$ = 0.6, 1.0, 1.8, 3.2, 5.6, and 10.0 (solid lines). 
Several examples of the thermal evolution obtained in the 3D hydrodynamic simulations 
(colored dashed lines with filled circles) are also plotted in this figure.
Our one-zone results can well explain the wide variety of thermal evolution. 
We briefly describe the essence of the evolution in each density range below. 

\paragraph{Low Density ($n_{\rm H,cen} < 10^8 \ {\rm cm^{-3}}$)} \

In this stage, HD cooling could significantly affect the thermal evolution 
for the slowly collapsing cloud.
HD cooling becomes efficient when the gas temperature falls below 100 K. 
The additional coolant HD further reduces the temperature to the cosmological temperature floor, $T_{\rm CMB}$.
Figure \ref{plot_100FS_one-zone3} shows that 
the gas temperature falls more strongly for the longer collapse timescale. 
The gas temperature falls below 100 K 
via ${\rm H_2}$ and HD cooling when $f > 1.8$. 
Indeed, this critical value of $f$ can be analytically derived by requiring 
the collapse timescale $t_{\rm coll} = f \cdot t_{\rm ff}$ 
to be equal to the ${\rm H_2}$ cooling time at T = 100 K 
(and $n_{\rm H} =  10^3 \ {\rm cm^{-3}}$). 
The resulting value of $f_{\rm crit} = 1.2$ is close to the numerical result.

\paragraph{Intermediate Density ($10^8 \ {\rm cm^{-3}} < n_{\rm H,cen} < 10^{12} \ {\rm cm^{-3}}$)} \

At densities above $10^8 \ {\rm cm^{-3}}$ 
the three-body ${\rm H_2}$ formation reaction becomes quite efficient, and
${\rm H_2}$ formation heating becomes more important than heating by compression.
Nevertheless, the equilibrium temperature is lower for cases
with a higher abundance of hydrogen molecules.
The middle panel of Figure \ref{plot_100FS_one-zone3} shows that 
the more slowly collapsing clouds have a higher ${\rm H_2}$ fraction 
at densities $10^8 \ {\rm cm^{-3}} $ $< n_{\rm H,cen} <$ $10^{12} \ {\rm cm^{-3}}$,
because more ${\rm H_2}$ is formed when the collapse time is longer.
${\rm H}_2$ cooling dominates the cooling, so the temperature is lower 
with the larger $f$ in this density range.

\paragraph{High Density ($n_{\rm H,cen} > 10^{12} \ {\rm cm^{-3}}$)} \

At these high densities all of the hydrogen has been converted
into its molecular form. Compression heating, which increases with decreasing 
collapse timescale, becomes the dominant heating process.
The heating rate is thus lower for the longer collapse timescale. 
This explains why the temperature remains lower 
with the larger $f$ in the highest density range.

\begin{figure}
\begin{center}
\resizebox{6cm}{!}{\includegraphics[clip]{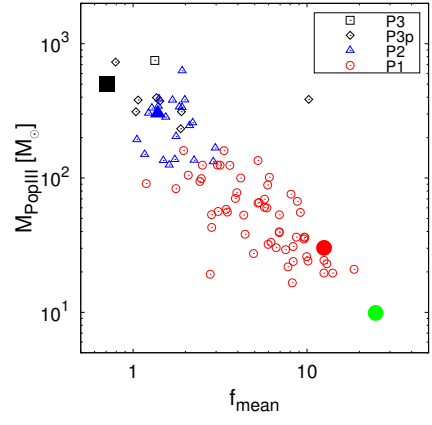}}
\caption{
Correlation between the ratio $f= t / t_{\rm ff}$ of the averaged relative collapse timescale to
the free-fall timescale and the final stellar masses in our numerical simulations. 
The different symbols denote the different paths of the protostellar evolution 
as in Figure \ref{plot_LIST_dmdt_multi2}.
}
\label{plot_100FS_fmean-MpopIII}
\end{center}
\end{figure}

\paragraph{} 
Figure \ref{plot_100FS_fmean-MpopIII} shows the correlation between 
$f$ and the final stellar masses in our numerical results. 
$f$ is evaluated in the following manner.
We calculate the timescale between two snapshots 
during which the central density decouples, $n_{\rm cen} \to 10 \times n_{\rm cen}$. 
$f$ is the ratio of this collapsing timescale, $t_{\rm coll}$, to the free-fall time, 
\begin{equation}
t_{\rm ff} = \left( \frac{3 \pi}{32 G \rho} \right)^{1/2} \ .
\end{equation}
We average these ratios at $n_{\rm cen} = 10^4$ to $10^{12} \ {\rm cm^{-3}}$ and 
assume it as the characteristic parameter of the collapsing cloud, $f$. 
The good correlation presented here suggests that 
the thermal evolution in our 3D cosmological simulations 
really depends on the parameter $f$.

\cite{mckee08} adopt the rotational degree of the cloud $f_{\rm Kepler}$ 
as a key parameter which determines 
the final stellar mass in their semi-analytic model. 
However, they only consider that the rotational support of 
the cloud slows down the mass accretion onto the star. 
We have shown that the rotational support reduces 
the accretion rate by changing the chemo-thermal evolution 
as well as the gas dynamics.

\end{document}